\documentclass[aps,prd,preprint,floatfix,nobibnotes,nofootinbib]{revtex4-2}

\usepackage{amsmath}
\usepackage{amssymb}
\usepackage{graphicx}
\usepackage{float}
\usepackage{dsfont}
\usepackage{ulem}
\usepackage[colorlinks=true]{hyperref}

\newcommand{\be}{\begin{equation}}
\newcommand{\ee}{\end{equation}}
\newcommand{\ba}{\begin{eqnarray}} 
\newcommand{\ea}{\end{eqnarray}} 
\newcommand{\nn}{\nonumber}

\newcommand{\bea}{\begin{eqnarray}}
\newcommand{\eea}{\end{eqnarray}}

\usepackage[usenames, dvipsnames]{xcolor}
\definecolor{darkgreen2}{RGB}{0, 150, 0}
\newcommand{\sm}{\textcolor{red}}

\numberwithin{equation}{section}

\usepackage{setspace}

\begin{document} 

\title{Angular momentum of glasma}

\author{Margaret E. Carrington}
\affiliation{Department of Physics, Brandon University,
Brandon, Manitoba R7A 6A9, Canada}
\affiliation{Winnipeg Institute for Theoretical Physics, Winnipeg, Manitoba, Canada}

\author{Stanis\l aw Mr\' owczy\' nski} 
\affiliation{National Centre for Nuclear Research, ul. Pasteura 7,  PL-02-093 Warsaw, Poland}

\date{June 12, 2026}

\begin{abstract}

The earliest phase of an ultrarelativistic heavy ion collision can be described as a highly populated system of gluons called glasma. We study some glasma characteristics related to the system's angular momentum. The first one is the global angular momentum perpendicular to the reaction plane, which is spanned by the beam axis and impact parameter vector. We show that only a small fraction of the enormous initial angular momentum of the colliding ultrarelativistic nuclei is transferred to the glasma.  Our main focus is the glasma angular momentum directed along the beam axis. This quantity has a local character and results from the inhomogeneous velocity field generated in the glasma. We calculate the vorticity and local angular momentum which show noticeably different behaviour. The results are analyzed in detail and discussed in the context of existing experimental data. We argue that neither vorticity nor thermal vorticity but instead the local angular momentum controls the polarization of final-state hadrons.

\end{abstract}

\maketitle
\newpage

%%%%%%%%%%%%%%%%%%%%%%%%%%%%%%%%%%%%%%%%%%%%%%%%%%%%%%%%
\section{Introduction}
\label{sec-intro}
%%%%%%%%%%%%%%%%%%%%%%%%%%%%%%%%%%%%%%%%%%%%%%%%%%%%%%%%

Observations of spin polarization phenomena in relativistic heavy-ion collisions opened new possibilities to study the collision dynamics and properties of the strongly interacting matter produced in these collisions. Global polarization of hyperons and vector mesons transverse to the reaction plane (defined as the plane spanned by the beam axis and impact parameter vector) was predicted \cite{Liang:2004ph,Voloshin:2004ha} and later discovered \cite{STAR:2017ckg,ALICE:2019aid,STAR:2022fan,ALICE:2022dyy}. A more subtle effect of spin polarization along the collision axis was conjectured \cite{Becattini:2017gcx,Voloshin:2017kqp} and then observed \cite{STAR:2019erd,STAR:2023eck,ALICE:2021pzu}. The experimental achievements reviewed in \cite{Niida:2024ntm} were accompanied by considerable theoretical activity aimed at a quantitative description of the newly discovered phenomena. The theoretical efforts are summarized in the review articles \cite{Florkowski:2018fap,Becattini:2024uha}.

The initial system of colliding nuclei carries angular momentum perpendicular to the reaction plane. When this angular momentum is transmitted to the quark-gluon plasma produced in the collision, it is responsible for the experimentally observed global spin polarization of hyperons at mid-rapidity. The situation with local polarization in the direction of the beam is much less clear. There is a simple argument that the polarization along the beam direction is generated by the vorticity of the strongly interacting fluid which is in turn caused by the gradient of the elliptic flow. The sign of the observed $\Lambda$ polarization contradicts this argument, which is known as the `spin sign puzzle'. Various versions of relativistic hydrodynamics have been formulated that take into account the spin degrees of freedom of the fluid components and use the idea that the spin polarization of final-state hadrons is driven by thermal vorticity, see \cite{Becattini:2024uha,Florkowski:2018fap}. The correct sign and magnitude can be obtained within the framework of hydrodynamic and statistical models, but one has to include, in addition to the thermal vorticity, what is called a shear effect \cite{Fu:2021pok,Becattini:2021iol,Alzhrani:2022dpi,Florkowski:2019voj}. Furthermore, these calculations are very sensitive to details of the collective flow and the precise formulation of the shear effect. For example, a hydrodynamic model  which includes both thermal vorticity and shear \cite{Yi:2024kwu} predicts the wrong sign of polarization for the very recent experimental data on the $\Lambda$ polarization along the beam axis in high-multiplicity p-Pb collisions at the LHC \cite{CMS:2025nqr}. 

In this paper we are interested in the very early state of the system that is far from thermodynamical equilibrium and preceeds the hydrodynamic phase. This earliest phase is a highly populated system of gluon fields called a glasma. The glasma is described by the Colour Glass Condensate effective theory (CGC), see the review articles \cite{Iancu:2003xm,Gelis:2010nm,Gelis:2012ri} and the original papers that pioneered the theory \cite{McLerran:1993ni,McLerran:1993ka,McLerran:1994vd,Kovner:1995ts,Kovner:1995ja,Dumitru:2001ux}.

The glasma phase is very anisotropic, the energy density reaches maximal values, and its non-Abelian dynamics is strongly non-linear. In a series of papers \cite{Carrington:2020ssh,Carrington:2021qvi,Carrington:2020sww,Carrington:2022bnv,Carrington:2021dvw,Carrington:2023nty,Carrington:2024utf} reviewed in \cite{Carrington:2024vpf} we have shown that various phenomena observed through characteristics of final states of heavy-ion collisions have their origin in the glasma phase. In particular, the experimentally observed pattern of collective flow quantified by the Fourier coefficients $v_1, v_2, v_3 \dots$ is already present in the glasma \cite{Carrington:2021qvi,Carrington:2023nty,Carrington:2024utf}. Gradients of the flow generate the vorticity which is commonly believed to be responsible for the spin polarization of final-state hadrons. 

We analyze the glasma global angular momentum perpendicular to the reaction plane. The quantity has already been studied in our earlier works \cite{Carrington:2021qvi,Carrington:2023nty} but we return to it here, paying particular attention to reducing the theoretical uncertainty of the result. The main topic of this paper is the local angular momentum of the glasma along the beam direction and the vorticity of the glasma fluid, with the velocity defined as the Poynting vector divided by the energy density. 

Throughout the paper we use the natural system of units with $c = \hbar = k_B =1$. Greek letters $\mu, \nu, \rho, \dots$ label components of four-vectors. Following common notation, two-component vectors transverse to the collision axis $z$ are denoted as $\vec x_\perp$ in Sec.~\ref{sec-method}, where our CGC calculations are described. In the following chapters, however, two-vectors are denoted as $\vec R$ or $\vec r$ without the subscript $\perp$ and their components are indexed with Latin letters $i,j,k, \dots$. Latin letters from the beginning of the alphabet $a,b, c \dots$ label colour components of elements of the SU($N_c$) gauge group in the adjoint representation. We use three systems of coordinates: Minkowski $(t, z, x, y)$, light-cone $(x^+,x^-, x, y)$ and Milne $(\tau, \eta, x, y)$, where $x^\pm = (t\pm z)/\sqrt{2}$, $\tau=\sqrt{t^2-z^2}$ and $\eta=\ln(x^+/x^-)/2$. The indices of vector and tensor components, like $A^t$, $F^{+x}$, $T^{\tau \eta}$, clearly show which coordinates are used.

%%%%%%%%%%%%%%%%%%%%%%%%%%%%%%%%%%%%%%%%%%%%%%%%%%%%%%%%
\section{Summary of the computational method}
\label{sec-method}
%%%%%%%%%%%%%%%%%%%%%%%%%%%%%%%%%%%%%%%%%%%%%%%%%%%%%%%%

The glasma angular momentum and vorticity are obtained from components of the energy-momentum tensor which is expressed in terms of chromodynamic fields and averaged over colour configurations of the colliding nuclei. To calculate the energy-momentum tensor we use a proper time expansion with a method introduced in \cite{Fries:2005yc} that is  designed to describe the earliest phase of relativistic heavy-ion collisions.  The technique was further developed in \cite{Fukushima:2007yk,Fujii:2008km,Chen:2015wia,Fries:2017ina}.  The procedure we use is described in detail in \cite{Carrington:2021qvi,Carrington:2023nty,Carrington:2024vpf}. In this series of papers we solved several technical problems with the method and demonstrated its ability to describe various physical characteristics of the glasma. The method is based on an expansion of the chromodynamic potentials in powers of the proper time, $\tau$. The proper time multiplied by the saturation scale $Q_s$ provides a dimensionless small parameter. This expansion allows one to solve the Yang-Mills equations iteratively. The results provided by the method are complicated in structure and limited to small values of $\tau$ but they are analytic and free of  numerical artifacts like those caused by taking a continuous limit in lattice calculations.

We summarize below the method to calculate the energy-momentum tensor of the glasma using a proper time expansion. More details can be found in \cite{Carrington:2021qvi,Carrington:2023nty,Carrington:2020ssh}. We consider a collision of two heavy ions moving with the speed of light towards each other along the $z$ axis and colliding at $t=z=0$. The vector potential of the gluon field is described with the ansatz \cite{Kovner:1995ts} 
\ba
\nn
A^+(x) &=& \Theta(x^+)\Theta(x^-) x^+ \alpha(\tau,\vec x_\perp) ,
\\\label{ansatz}
A^-(x) &=& -\Theta(x^+)\Theta(x^-) x^- \alpha(\tau,\vec x_\perp) ,
\\ \nn
A^i(x) &=& \Theta(x^+)\Theta(x^-) \alpha_\perp^i(\tau,\vec x_\perp)
+\Theta(-x^+)\Theta(x^-) \beta_1^i(x^-,\vec x_\perp)
+\Theta(x^+)\Theta(-x^-) \beta_2^i(x^+,\vec x_\perp) ,
\ea
where the functions $\beta_1^i(x^-,\vec x_\perp)$ and $\beta_2^i(x^+,\vec x_\perp)$ represent the pre-collision potentials, and the functions $\alpha(\tau,\vec x_\perp)$ and $\alpha_\perp^i(\tau,\vec x_\perp)$ are the post-collision potentials. 
In the forward light-cone the vector potential satisfies the sourceless Yang-Mills equations but the sources enter through boundary conditions that connect the pre-collision and post-collision potentials. The boundary conditions are
\ba
\label{cond1}
\alpha^{i}_\perp(0,\vec{x}_\perp) &=& \alpha^{i(0)}_\perp(\vec{x}_\perp) 
= \lim_{\text{w}\to 0}\left(\beta^i_1 (x^-,\vec{x}_\perp) + \beta^i_2(x^+,\vec{x}_\perp)\right) ,
\\
\label{cond2}
\alpha(0,\vec{x}_\perp) &=& \alpha^{(0)}(\vec{x}_\perp) 
= -\frac{ig}{2}\lim_{\text{w}\to 0}\;[\beta^i_1 (x^-,\vec{x}_\perp),\beta^i_2 (x^+,\vec{x}_\perp)] ,
\ea
where $x^{\pm} \in [- \text{w}/2, \text{w}/2]$ and the notation $\lim_{\text{w}\to 0}$ indicates that the width of the sources across the light-cone is taken to zero, as the colliding nuclei are infinitely contracted. 

We find solutions valid for early post-collision times by expanding the Yang-Mills equations in the proper time $\tau$. Using these solutions we can write the post-collision field-strength tensor, and energy-momentum tensor, in terms of the initial potentials $\alpha(0, \vec x_\perp)$ and $\vec\alpha_\perp(0, \vec x_\perp)$ and their derivatives, which in turn are expressed through the pre-collision potentials $\vec \beta_1(x^-,\vec x_\perp)$ and $\vec \beta_2(x^+,\vec x_\perp)$ and their derivatives. 

The next step is to use the Yang-Mills equations to write the pre-collision potentials in terms of the colour charge distributions of the incoming ions. One then averages over a Gaussian distribution of colour charges within each nucleus. The average of a product of colour charges can be written as a sum of terms that are products of the averages of all possible pairs, which is called Wick's theorem. We use the Glasma Graph approximation \cite{Lappi:2017skr} which means that we apply Wick's theorem not to colour charges but to gauge potentials.

The correlator of two pre-collision potentials from different ions is assumed to be zero as the potentials are not correlated to each other. All physical quantities we study are constructed from the correlators for two potentials from the same ion
\be
\label{core5-20}
\delta^{ab} B_n^{ij}(\vec{x}_\perp,\vec y_\perp) \equiv 
\lim_{{\rm w} \to 0}  \langle \beta_{n\,a}^i(x^-,\vec x_\perp) \beta_{n\,b}^j(y^-,\vec y_\perp)\rangle  ,
~~~~~~n=1,~\,2
\ee
and their derivatives. The correlators (\ref{core5-20}) are expressed through the colour charge surface density of a given ion $\mu_1(\vec x_\perp)$ or $\mu_2(\vec x_\perp)$, see Sec.~II of Ref.~\cite{Carrington:2021qvi}, which is assumed to be proportional to the nuclear density of the ion, as in the IP-Glasma model \cite{Schenke:2012wb,Schenke:2012hg,Schenke:2020mbo,Mantysaari:2025kls}. In our calculation we take the colour charge surface density to be a two-dimensional projection of a Woods-Saxon distribution onto the plane transverse to the collision axis
\be
\label{def-mu2-2}
\mu(\vec x_\perp)  =  
\Big(\frac{A}{207}\Big)^{1/3}\frac{\bar\mu}{2a\ln(1+e^{R_A/a})} 
\int^\infty_{-\infty}\frac{dz}{1 + \exp\big[(\sqrt{\vec x_\perp^2 + z^2} - R_A)/a\big]} \,.
\ee
The parameters $R_A$ and $a$ give the radius and skin thickness of a nucleus of mass number $A$. We use $r_0=1.25$ fm and $a=0.5$ fm so that the radius of a nucleus with $A=207$ is $R_A=r_0 A^{1/3}= 7.4$ fm. We  use the relation $\bar\mu = Q_s^2/g^4$ where $Q_s$ is the saturation scale and $\bar\mu$ is the value of $\mu(\vec x_\perp)$ at the center of the nucleus. While there are different conventions in the literature for the equations that relate the colour charge density $\rho$ to $\bar\mu$, and $\bar\mu$ to $Q_s$, the relation $\langle \rho\rho\rangle\sim Q^2_s/g^2$ contains the fundamental physics of the CGC approach and is consistently of the same form in the literature. Our definitions agrees with that of Refs.~\cite{Chen:2015wia,Lappi:2007ku,Schenke:2012wb,Epelbaum:2013ekf} after adjusting for changes in the notation. The proportionality factor in this relation cannot be determined within the CGC approach that we use, and we have made the conventional choice, see \cite{Chen:2015wia,Epelbaum:2013ekf} and set it to one. The glasma energy-momentum tensor will depend on this factor to some extent, but quantities determined from ratios of components of the energy-momentum tensor (like the ratio of two moments of the angular momentum field $L^z$ (see Eq.~(\ref{L-mom-def})) will depend on it more weakly. In Ref.~\cite{Lappi:2007ku} it is argued more generally that the dependence of physical quantities on this factor is weak. 

All of the results presented in this paper are obtained for the SU(3) gauge group, the saturation scale $Q_s = 2$ GeV, and infrared cutoff $m=0.2$ GeV. The infrared scale is identified with the inverse nucleon radius or the confinement scale and the value 0.2~GeV is standard. The choice $Q_s=2.0$~GeV is commonly used in similar calculations (where $Q_s$ is taken to be constant), see for example \cite{Epelbaum:2013ekf,Iancu:2002tr,Gelfand:2016yho}. In our earlier works \cite{Carrington:2021qvi,Carrington:2022bnv} we showed that the pressure anistropy, energy loss and momentum broadening coefficient are fairly insensitive to these values. 

In our calculations we set the coupling constant $g$ to one. The reason is explained below. It is well known that in classical gluodynamics the coupling constant can be scaled out by means of the redefinition $A_a^\mu \to \hat A_a^\mu/g$. After rescaling the Lagrangian density depends on $g$ only through the prefactor $g^{-2}$ and the coupling constant does not enter the Yang-Mills equations or their solutions. The CGC approach that we use is a classical effective theory that depends on the infrared and ultraviolet cut-offs $m$ and $Q_s$ and it is not obvious that this scaling property survives in the effective theory. 
The two-point correlator of two pre-collision potentials, denoted $B$, depends on $g$ and $\bar{\mu}$ only through the product $g^2 \bar{\mu}$ (see  Eq.~(43) of Ref.~24) and since the CGC assumption is $\bar{\mu} \sim Q_s^2/g^4$ we have (supressing all indices) $B = \langle\beta\beta\rangle \sim g^2 \bar{\mu}\sim Q_s^2/g^2$. The correlator of two scaled pre-collision potentials (defined $\hat\beta=g\beta$) is therefore independent of $g$. The energy-momentum tensor is constructed from these two-point correlators and therefore it depends on the coupling only through the prefactor $g^{-2}$. If we use ${\cal C}$ to denote the unknown proportionality factor in the relation $\bar \mu \sim Q_s^4/g^4$ (see above) then we have an overall factor ${\cal C}^2/g^2$ that multiplies our energy-momentum tensor. In our calculation we set ${\cal C}=g=1$ for simplicity (it would obviously be trivial to rescale the energy-momentum tensor by $1/g^2$ for any value of $g$). The consequence is that any quantity that depends linearly on the energy-momentum tensor can only be considered an order of magnitude estimate. 

If not stated otherwise the glasma energy-momentum tensor is calculated for Pb-Pb collisions at eighth order of the proper time expansion (all contributions up to eighth order are summed), and all calculations are done at  mid-rapidity ($\eta=z=0$) and at $\tau = t = 0.06$~fm. All distances are given in femtometer (fm). 

\begin{figure}[t]
\centering
\includegraphics[width=10cm]{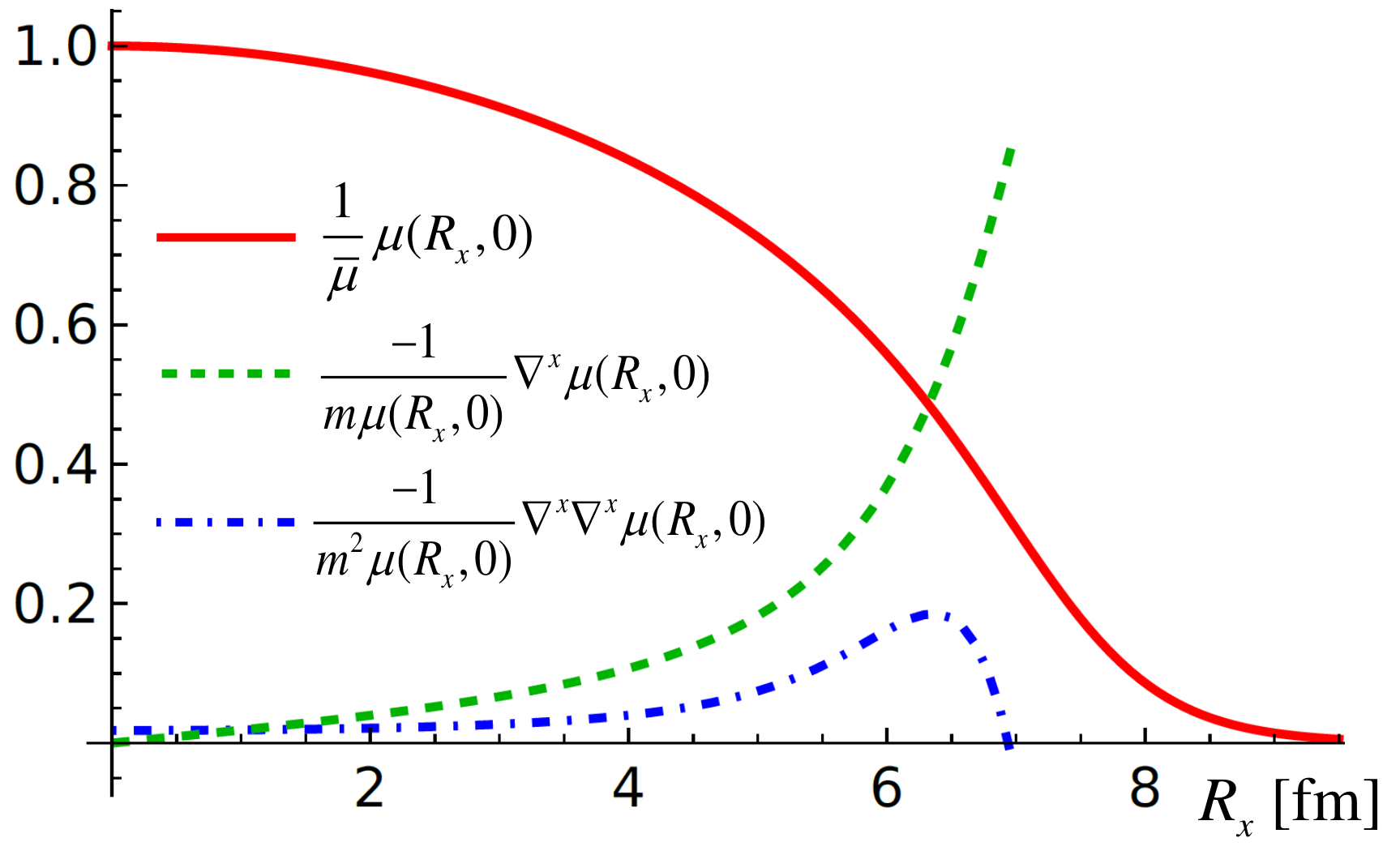}
\vspace{-2mm}
\caption{The red (solid), green (dashed) and blue (dot-dashed) curves show the colour charge density and its first and second derivatives as a function of distance from the center of the nucleus.
\label{MUd-plot}}
\end{figure}

To obtain analytic results for the correlators in Eq.~(\ref{core5-20}) we perform a gradient expansion. We define the vector $\vec {\cal R} = (\vec x_\perp + \vec y_\perp)/2$ and expand the colour charge densities (\ref{def-mu2-2}) around the centers of colliding nuclei and we take into account only the first two terms in the expansion. Figure~\ref{MUd-plot} shows $\mu({\cal R}_x,0)$ and its derivatives as functions of the distance from the center of the charge distribution.  The figure shows that the gradient expansion can be trusted for ${\cal R}\lesssim {\cal R}_{\rm max} = 6.5$~fm, where the magnitude of the derivatives of the charge density are less than about 50\% of the charge density itself. 

To consider collisions with nonzero impact parameter we displace the centers of the two ions by $\pm \vec b/2$ where the impact parameter vector $\vec b$ is directed along the $x$-axis. To do this we use $\mu_1(\vec{\cal R}) \equiv \mu(\vec R-\vec b/2)$ and $\mu_2(\vec{\cal R}) \equiv \mu(\vec R + \vec b/2)$. The trajectory of the first nucleus, which moves along the $z$~axis, is at $x=b/2$ and the second nucleus, which moves in the negative $z$~direction, has its center at $x=-b/2$. The position $\vec R=0$ is the center of the interaction region. We note that since $\vec{\cal R}=\vec{R}\pm \vec{b}/2$ a fixed upper bound ${\cal R}_{\rm max}$ (see Fig.~\ref{MUd-plot}) will correspond to different values of $R_{\rm max}$ depending on the impact parameter. It is therefore important to be very careful about how we handle the gradient expansion in calculations that compare any quantity at different values of impact parameter. We also point out that calculations of angular momentum are particularly sensitive to the gradient expansion. Physically we know that the dominant contribution to the global angular momentum of the system will come from the edges of the system, where the material is farthest from its center. From Fig.~\ref{MUd-plot} it is clear that this is the domain where the gradient expansion is least trustworthy. 

\begin{figure}[t]
\begin{center}
\includegraphics[width=16cm]{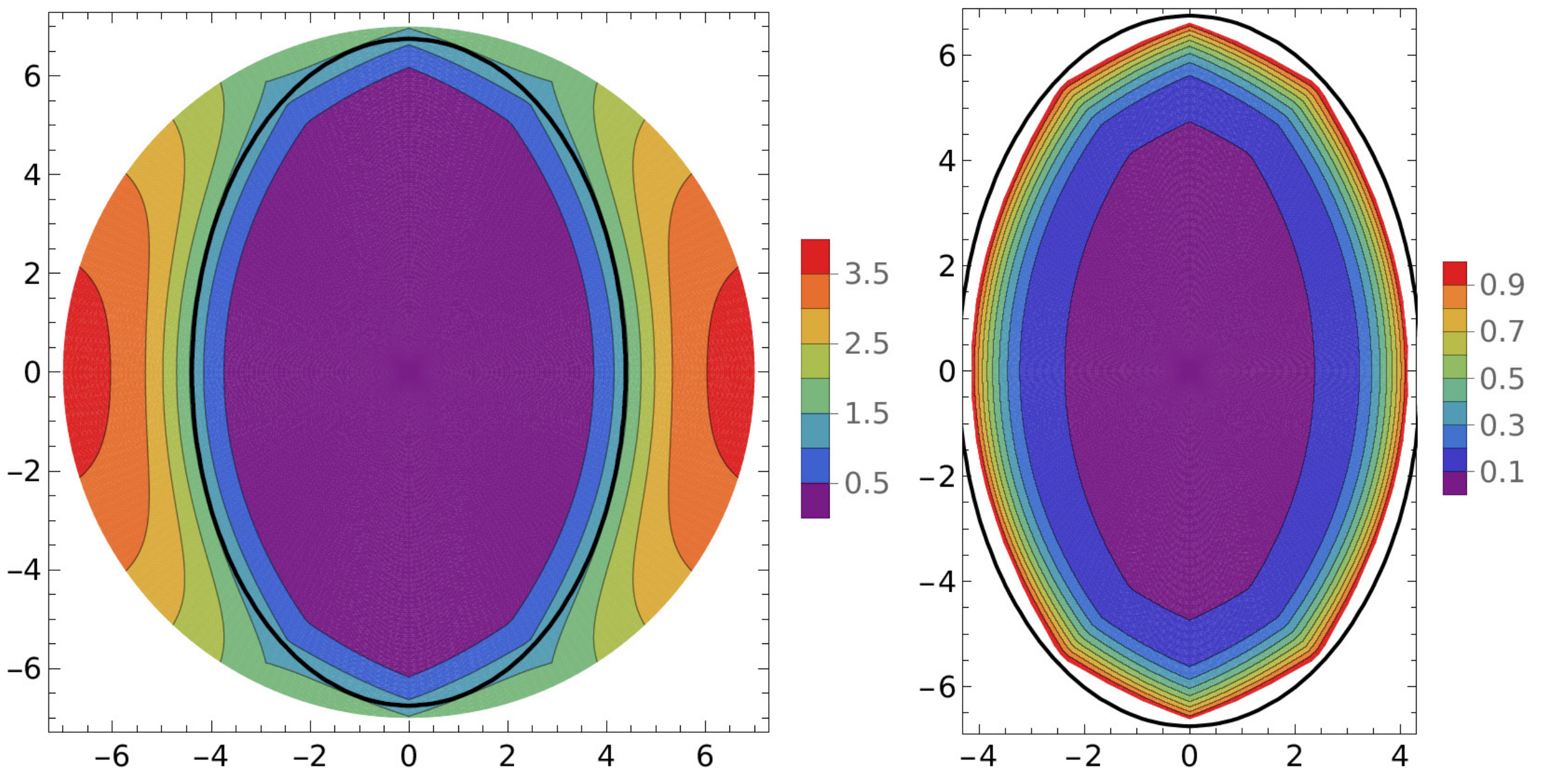}
\end{center}
\vspace{-7mm}
\caption{The value of $\delta(R_x,R_y)$ inside a circle of radius $7$~fm (left panel) and the region for which $\delta(R_x,R_y)<1$ (right panel), for a collision with impact parameter $b=6$~fm. The axes show $R_x$ and $R_y$ in fm.}
\label{fig-ellipse-2} 
\end{figure}

To define the parameter that must be small for the gradient expansion to be valid for a collision with nonzero impact parameter we define
\be
\mu_n^{(i,j)}(R_x,R_y) \equiv \frac{d^i}{dR_x^i}\frac{d^j}{dR_y^j}\mu_n(R_x,R_y) ,
~~~~~~~~~~
\delta_n^{(i,j)}(R_x,R_y) \equiv \frac{\mu_n^{(i,j)}(R_x,R_y)}{m^{i+j} \mu_n(R_x,R_y)}
\ee
and consider $[\delta_n^{(i,j)}]^2$ for $(i,j)\in\{(1,0),(0,1)\}$ and $\delta_n^{(i,j)}$ for $(i,j)\in\{(2,0),(0,2),(1,1)\}$, where $n\in\{1,2\}$ to include both surface densities $\mu_1$ and $\mu_2$. We define $\delta(R_x,R_y)$ as the maximum value of these 10 numbers. The region where we trust the gradient expansion is the part of the transverse plane where $\delta(R_x,R_y)< \bar\delta$ with $\bar\delta<1$. In our calculations we use two different methods to implement this constraint. We explain them below. 

\begin{figure}[t]
\begin{center}
\includegraphics[width=12cm]{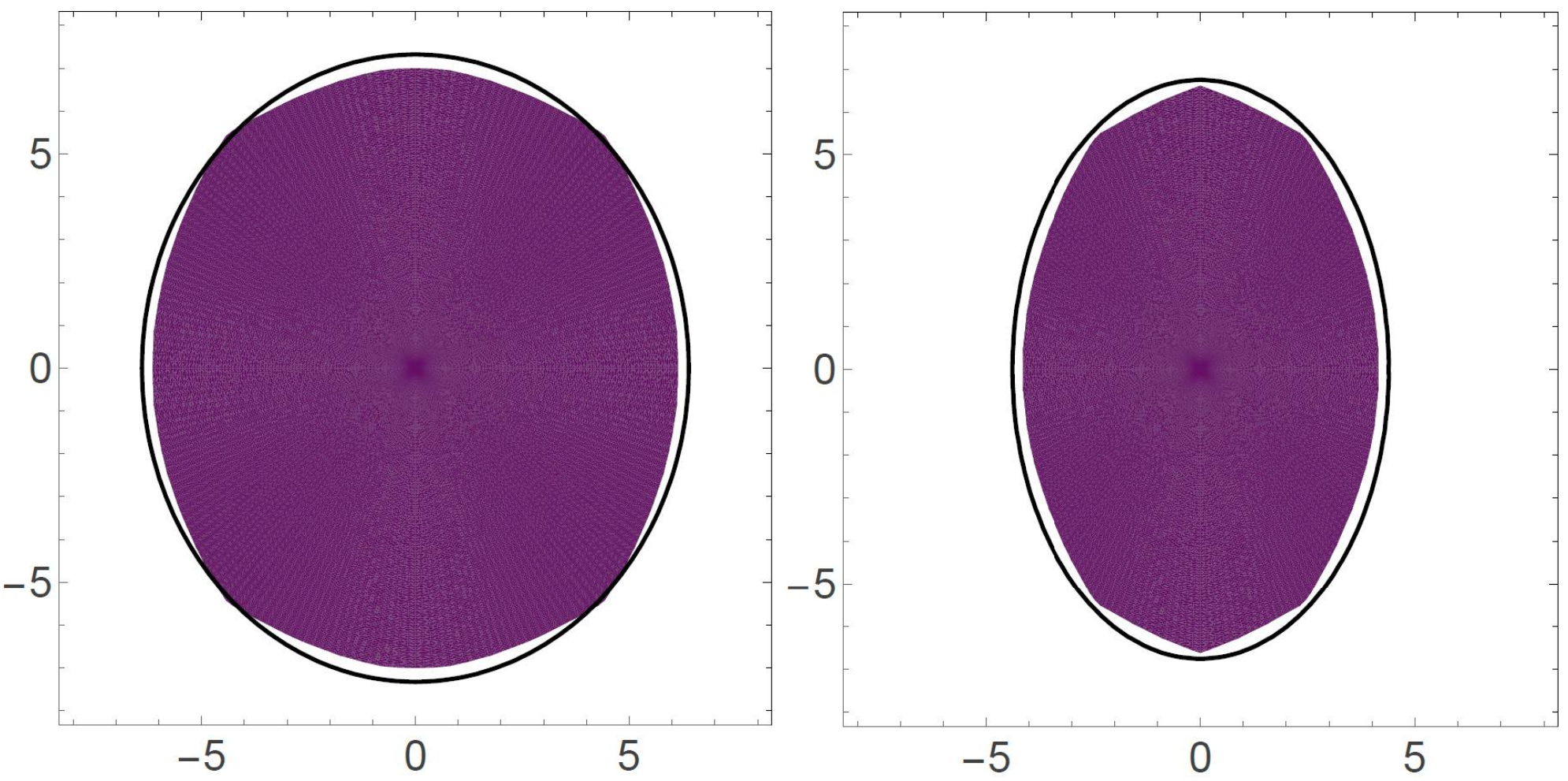}
\end{center}
\vspace{-7mm}
\caption{The part of the transverse plane for which $\delta(R_x,R_y)<1$ and the elliptical interaction region  (outlined by the black curve) for $b=2$~fm (left) and $b=6$~fm (right).}
\label{fig-ellipse} 
\end{figure}

We define the interaction region as the overlap of two circles of radius $R_A$ and centers at $(\pm b/2,0)$. Since this almond shaped region and an ellipse with semi minor/major axes given by ${\cal A}=R_A-b/2$ and ${\cal B}=\sqrt{R_A^2- b^2/4}$ are almost identical, we use an ellipse for simplicity in some calculations. In the left panel of Fig.~\ref{fig-ellipse-2} we show the values of $\delta(R_x,R_y)$ in a circle of radius $7$~fm for a collision with impact parameter $b=6$~fm. In the right panel we show only the region where $\delta(R_x,R_y)<1$. In both cases the black curve indicates the interaction region. The figure shows that inside the interaction region the gradient expansion that we use is valid. 
Figure~\ref{fig-ellipse} shows that for both large and small impact parameters there is good agreement between the interaction region and the part of the transverse plane for which $\delta(R_x,R_y) <1.0$.

In most of the calculations in this paper, when we take an integral over the transverse plane, we include only points for which the constraint $\delta(R_x,R_y)<0.9$ is satisfied. The value $\delta(R_x,R_y)=0.9$  is fairly large compared to typical choices for a parameter that characterises a perturbative expansion.  It is very natural to think that reducing the value of $\delta(R_x,R_y)$ will produce more accurate results. We emphasize that in our calculations this is not true. The point is that if we use a very small value of $\delta(R_x,R_y)$ we will get a very accurate result for the quantity we are calculating, but that quantity does not actually represent the angular momentum of the glasma, because it does not include the physically important part of the transverse plane. Our choice for the value of $\delta(R_x,R_y)$ is justified by two facts. Firstly, there is physical motivation which is that the region selected by the condition $\delta(R_x,R_y)<0.9$ corresponds well with the interaction region, which is the part of the transverse plane occupied by the glasma at the earliest stage of the collision. The matching of the interaction region and the part of the transverse plane where $\delta(R_x,R_y)$ is small is illustrated in Figs.~\ref{fig-ellipse-2} and \ref{fig-ellipse}. Secondly, our results are largely independent of the choice of $\bar\delta$, as verified by our error analysis in  Secs.~\ref{Ly-sec} and \ref{Lz-sec}.

%%%%%%%%%%%%%%%%%%%%%%%%%%%%%%%%%%%%%%%%%%%%%%%%%%%%%%%%
\section{Formulas for angular momentum}
%%%%%%%%%%%%%%%%%%%%%%%%%%%%%%%%%%%%%%%%%%%%%%%%%%%%%%%%

In our previous work \cite{Carrington:2021qvi} we derived an expression for the angular momentum of the glasma per unit rapidity\footnote{Our method is similar to that of Ref.~\cite{Fries:2017ina} where a slightly inconsistent procedure is used. In that paper the authors define angular momentum in Minkowski space on a surface of constant time, and then enforce separately that the integral should be calculated with $\tau$ held fixed.}. We repeat the basic steps of this calculation below. We define the tensor
\be
\label{M-def}
M^{\mu\nu\rho}=T^{\mu\nu}R^\rho - T^{\mu\rho}R^\nu ,
\ee
where $R^\mu$ denotes a component of the position vector. The energy-momentum tensor is divergenceless and therefore the tensor (\ref{M-def}) satisfies the equation $\nabla_\mu M^{\mu\nu\rho} = 0$. Using Stokes' theorem one obtains a set of six conserved quantities
\be
\label{J1}
J^{\nu\rho} = \int_\Sigma d^3 y \sqrt{|\gamma|} \, n_\mu M^{\mu\nu\rho} ,
\ee
where $n^\mu$ is a unit vector perpendicular to the hypersurface $\Sigma$, $\gamma$ is the induced metric on this hypersurface, and $d^3y$ is the corresponding volume element. 

The angular momentum is obtained from the Pauli-Lubanski four-vector 
\be
\label{L-def}
L_\mu = -\frac{1}{2}\epsilon_{\mu\nu\rho\sigma}J^{\nu\rho}U^\sigma ,
\ee
where $U^\sigma$ is the vector that denotes the rest frame of the system. We work in Milne coordinates and use $n_\mu = (1,0,0,0)$ so that the tensor $J^{\nu\rho}$ is defined on a hypersurface of constant proper time $\tau$ as
\be
\label{J-now}
J^{\nu\rho} = \tau \int d\eta \int d^2 R \, M^{0\nu\rho} .
\ee
In the rest frame, where $U^\gamma = (1,0,0,0)$, one finds
\be
\label{L-def-2}
L_\mu = \frac{1}{2}\tau \, \epsilon_{0\mu\nu\rho} 
\int d\eta \int d^2 R \, \big(T^{0\nu}R^\rho - T^{0\rho}R^\nu \big) .
\ee
From Eq.~(\ref{L-def-2}) we get the angular momentum per unit rapidity 
\be
\label{L-def-3}
\frac{dL_\mu}{d\eta} = \frac{1}{2}\tau \, \epsilon_{0\mu\nu\rho} 
\int d^2 R \, \big(T^{0\nu}R^\rho - T^{0\rho}R^\nu \big) .
\ee

The energy-momentum tensor in the formula (\ref{L-def-3}) is expressed through the classical chromodynamic vector fields. We use the Belinfante improved form of the energy-momentum tensor which is symmetric and gauge invariant (see Eq.~(14) of our earlier work \cite{Carrington:2020ssh}). The angular momentum obtained from the Pauli-Lubanski vector (\ref{L-def-3}) includes both orbital angular momentum and spin, which cannot be separated from each other in a gauge invariant way \cite{Ji:1996ek}. 

We are interested in the $y$ and $z$~components of the angular momentum per unity rapidity in Minkowski space at mid-rapidity which are of the expected form
\ba
\label{Ly-result}
\frac{dL^y}{d\eta} &=& -\tau \int d^2 R \, R^x T^{0z} ,
\\ [2mm]
\label{Lz-result}
\frac{dL^z}{d\eta} &=& -\tau \int d^2 R \, \big(R^y T^{0x} - R^x T^{0y} \big). 
\ea
The formula (\ref{Ly-result}) will be used to calculate the glasma global angular momentum calculated around the center of the interaction region. For this purpose the integral will be performed over the  transverse cross-section of the volume occupied by the glasma including all points that satisfy the constraint $\delta(R_x,R_y)<\bar\delta$ with $\bar\delta$ typically set to $0.9$. The formula (\ref{Lz-result}) will provide the glasma angular momentum along the beam axis. Due to the system's symmetry the integral vanishes  if calculated over the entire transverse plane. Therefore the integral will be taken over a large set of small domains to obtain the local angular momentum. In these calculations the vector $\vec R$ is shifted so that the angular momentum is calculated around the center of each region.

%%%%%%%%%%%%%%%%%%%%%%%%%%%%%%%%%%%%%%%%%%%%%%%%%%%%%%%%
\section{Global angular momentum}
\label{Ly-sec}
%%%%%%%%%%%%%%%%%%%%%%%%%%%%%%%%%%%%%%%%%%%%%%%%%%%%%%%%

The glasma is expected to have significant global angular momentum inherited from the enormous angular momentum of the incoming nuclei when they collide at a nonzero impact parameter. The angular momentum carried by the nucleons which will participate in the collision is of order $10^5$ at maximum RHIC energies \cite{Gao:2007bc,Becattini:2007sr} and even larger at LHC energies. 

The results of our calculation for the global angular momentum of the glasma are shown below in Figs.~\ref{fig-Ly-vs-b} and \ref{fig-Ly-vs-t}. 
As explained at the end of Sec.~\ref{sec-method}, the calculation is sensitive to the value of the parameter $\bar\delta$ that characterises the validity of the gradient expansion. It is therefore crucial to test the extent to which our results are independent of the choice we make for $\bar\delta$. 
At the end of this section we discuss the procedure we use to do this.  

In Fig.~\ref{fig-Ly-vs-b} we show $-dL^y/d\eta$ versus impact parameter for different values of $\tau$ and $A$. The figure shows the results at sixth and eigth order in the proper time expansion, to verify convergence at the chosen values of $\tau$. The global angular momentum is negative because the collision geometry we use (the ion moving in the positive $z$~direction has its center displaced in the positive $x$~direction, and the ion moving in the opposite direction is displaced the same distance in the negative $x$~direction) produces angular momentum antiparallel to the $y$~axis. One also sees that the angular momentum initially increases with $b$, as expected, and then starts to drop off as the ions become very widely separated and the interaction region decreases. 

\begin{figure}[t]
\begin{center}
\includegraphics[width=10cm]{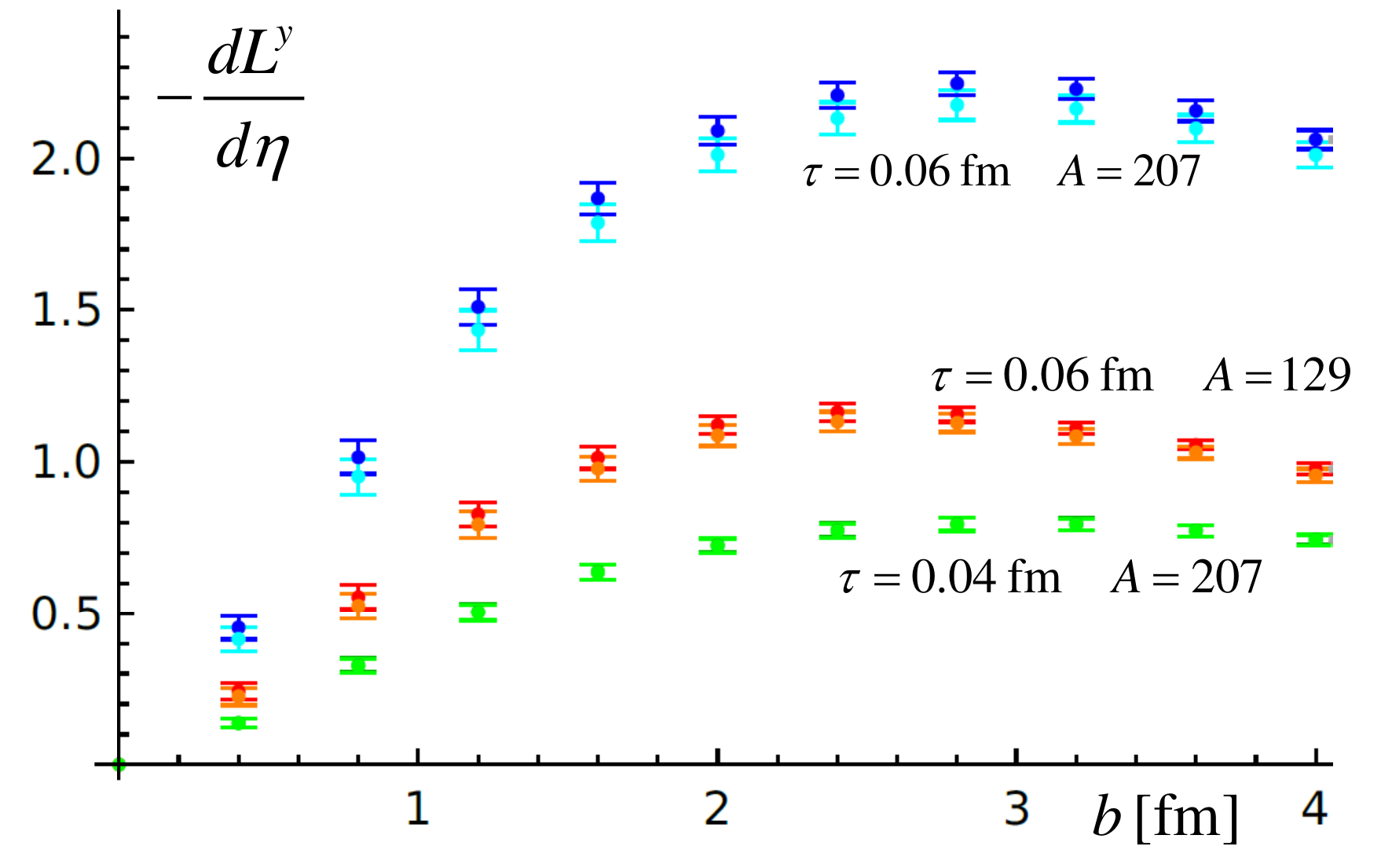}
\end{center}
\vspace{-8mm}
\caption{$-dL^y/d\eta$ versus $b$ for different values of $\tau$ and $A$. The lighter curves show the result at sixth order in the proper time expansion and the darker colours correspond to the eighth order calculation. }
\label{fig-Ly-vs-b}
\end{figure}

\begin{figure}[b]
\begin{center}
\includegraphics[width=10cm]{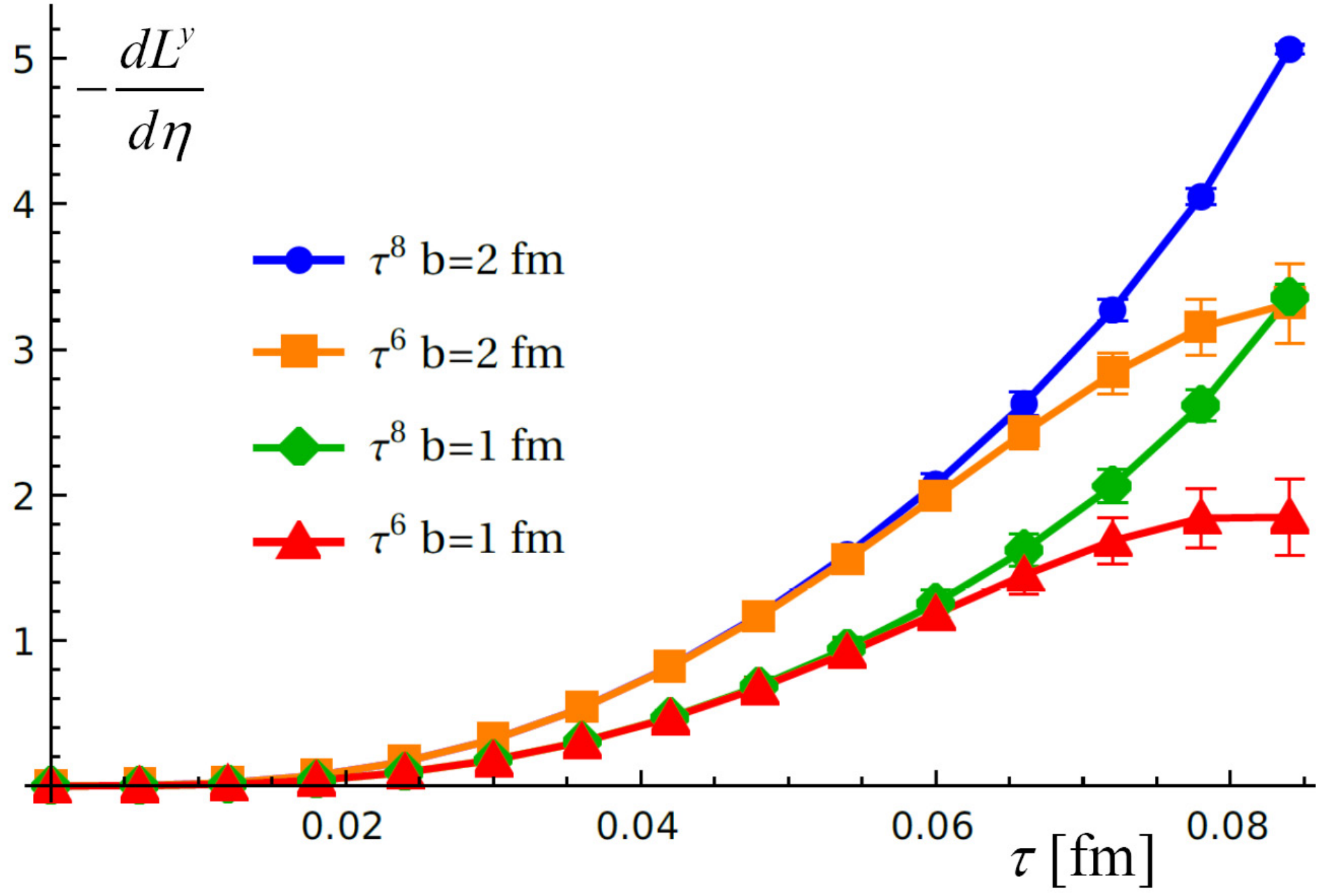}
\end{center}
\vspace{-8mm}
\caption{$-dL^y/d\eta$ versus $\tau$ for impact parameter $b=1$~fm and  $b=2$~fm.}
\label{fig-Ly-vs-t}
\end{figure}

In Fig.~\ref{fig-Ly-vs-t} we show the angular momentum versus proper time for two different impact parameters at sixth and eighth order in the proper time expansion with $A=207$. The sixth and eighth order results  agree with each other very well for times smaller than about 0.06 fm. For longer times the curves obtained at the fourth and eighth orders show rapid growth. The second and sixth order results suggest the saturation of $-dL^y/d\eta$. This is expected on  physical grounds because the angular momentum is transferred from the incoming nuclei to the glasma only during the short period when the nuclei pass through each other. 

The results in Figs.~\ref{fig-Ly-vs-b} and \ref{fig-Ly-vs-t} depend on the value of $\bar\delta$ that is used to do the calculation. We discuss below how to choose a physically motivated value, and how to verify that results are not strongly controlled by this choice. The chosen value of $\bar\delta$ must be small enough for the gradient expansion to be valid, but also large enough that the surface integral in Eq.~(\ref{Ly-result}) coincides with the interaction region which is physically important. Because of the competing nature of these two effects, the numerical result for $-dL^y/d\eta$ does not change monotonically with $\bar\delta$. As $\bar\delta$ increases from zero, $-dL^y/d\eta$ increases because more  of the interaction region is included, but at some point (with a value of $\bar\delta$ that depends on impact parameter) $-dL^y/d\eta$ starts to decrease, because the gradient expansion fails and produces a result which is too small. The obtain the results shown in Figs.~\ref{fig-Ly-vs-b} and \ref{fig-Ly-vs-t} we use a set of 51 values of $\bar\delta$ between 0.75 and 0.95, and the average and standard deviation of these calculations gives the value of $-dL^y/d\eta$ and its error. 

\begin{figure}[t]
\begin{center}
\includegraphics[width=9.5cm]{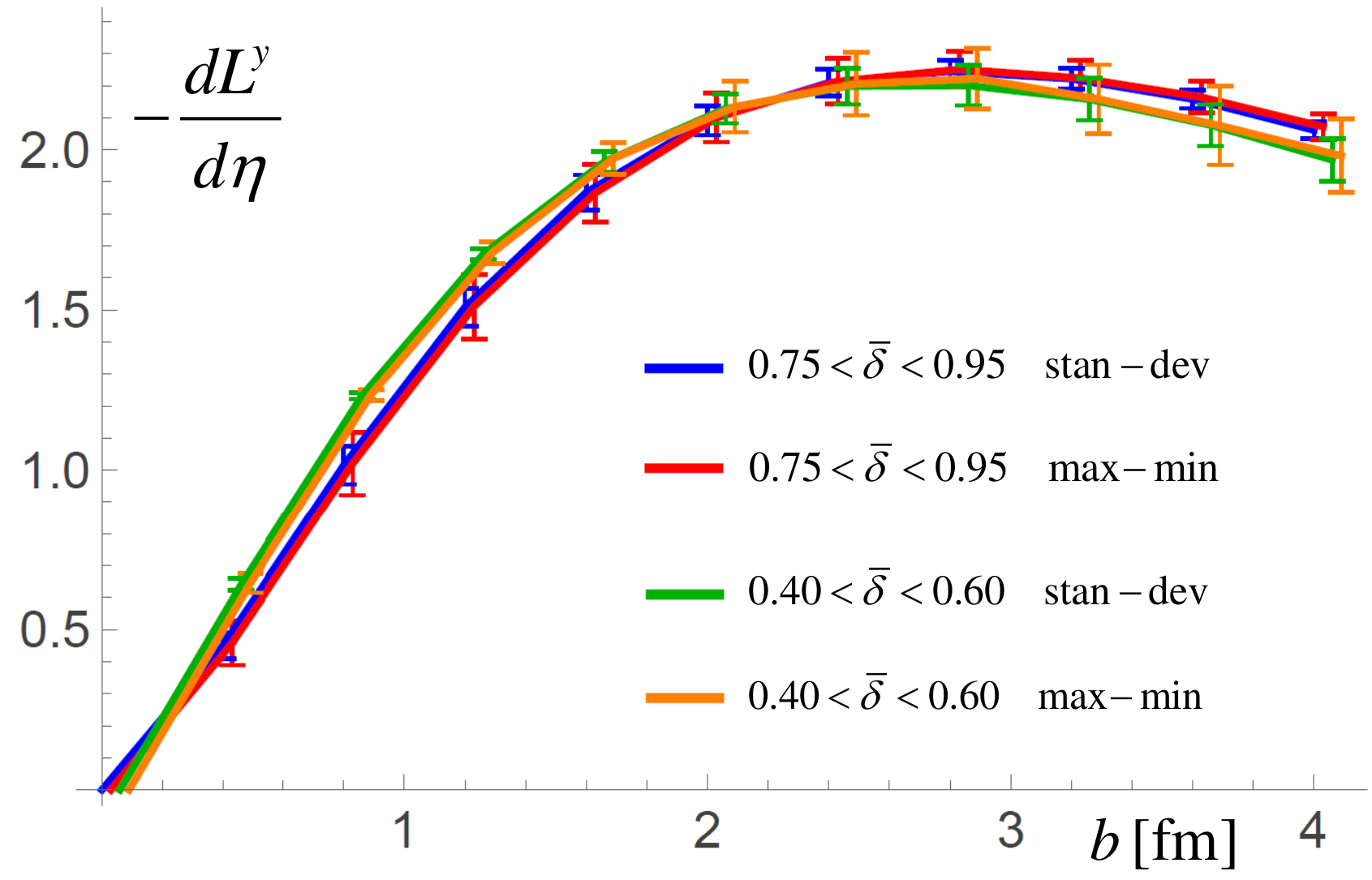}
\end{center}
\vspace{-8mm}
\caption{$-dL^y/d\eta$ versus $b$ for $A=207$ and $\tau=0.06$~fm using different methods to calculate the error and different ranges of values $\bar\delta$. The $x$-coordinates ($b$~values) of the curves are slightly shifted relative to the original (the dark blue curve) to make the error bars easier to distinguish.}
\label{fig-Ly-vs-b-2}
\end{figure}

To verify that this procedure gives an accurate result and a reliable estimation of its error, we have recalculated the result shown in the dark blue curve in Fig.~\ref{fig-Ly-vs-b} using three different methods. First we repeat the calculation using $\bar\delta \in(0.4,0.6)$ instead of $\bar\delta  \in(0.75,0.95)$, which should slightly suppress the result at large impact parameter because part of the interaction region is excluded. 
Secondly, using the same two ranges of $\bar\delta$ values, we replace the standard deviation with a worst case estimate defined as $\frac{1}{2}\left(|\text{max-ave}|+|\text{min-ave}|\right)$, or one half of the absolute value of the difference between the maximum and the average plus the absolute value of the difference between the minimum and the average. The results of these four calculations are seen in Fig.~\ref{fig-Ly-vs-b-2} which shows that both the peak and shape of the curve is largely independent of both the range of $\bar\delta$ values and the method used to calculate the error. We also note that the results for $-dL^y/d\eta$ presented here fully agree with those from our earlier works \cite{Carrington:2021qvi,Carrington:2023nty} but are more accurate and reliable because of the improvements in our error analysis procedure described above. 

The main point of this section is that our results, shown in figures~\ref{fig-Ly-vs-b}  and \ref{fig-Ly-vs-t}, indicate that the glasma carries only a very small imprint of the primordial angular momentum, which means that the majority of the angular momentum is carried by valence quarks. This result casts doubt on the idea that ultrarelativistic heavy-ion collisions produce a rapidly rotating quark-gluon plasma. We note that a recent work \cite{Akridge:2025jgy} found that the angular momentum transferred to the matter produced in heavy-ion collisions is significantly larger. However, these calculations were focussed on  lower collisions energies and it is unclear to what extent an extrapolation to the energy $\sqrt{s_{NN}}=500$~GeV is reliable.

%%%%%%%%%%%%%%%%%%%%%%%%%%%%%%%%%%%%%%%%%%%%%%%%%%%%%%%%
\section{Angular momentum along the beam direction}
\label{Lz-sec}
%%%%%%%%%%%%%%%%%%%%%%%%%%%%%%%%%%%%%%%%%%%%%%%%%%%%%%%%

\begin{figure}[t]
\begin{center}
\includegraphics[width=12cm]{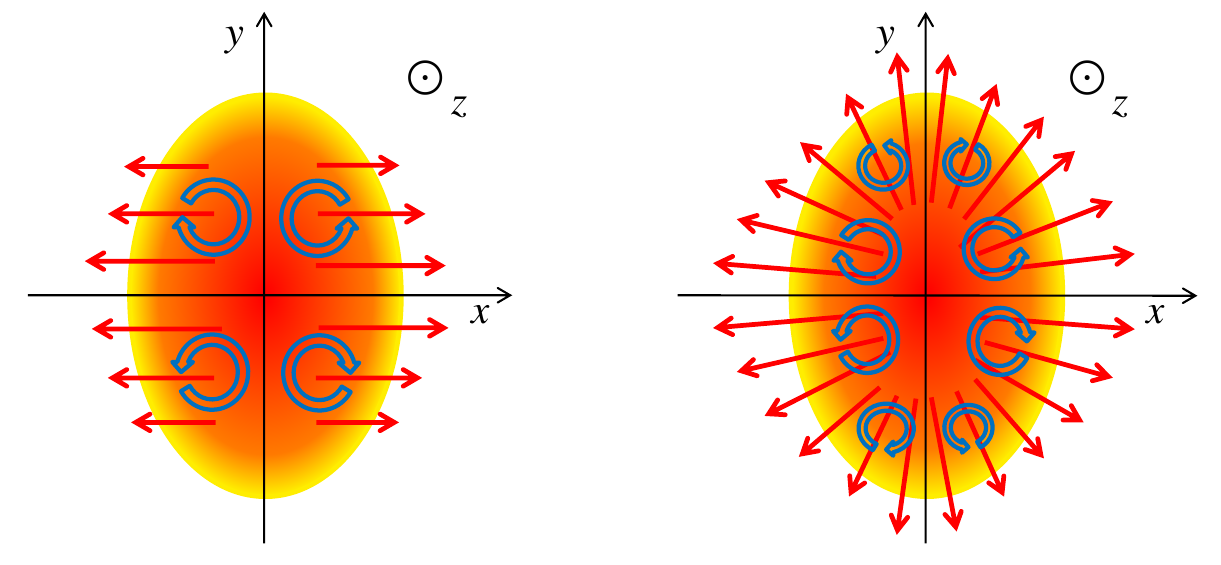}
\end{center}
\vspace{-7mm}
\caption{Elliptic flow generates vorticity. The red arrows represent the collective flow and the double blue arrows show the vorticity.}
 \label{fig-argument} 
\end{figure}

As explained in the introduction, our main motivation is to understand the origin of the local polarization in the direction of the beam. There is a simple argument \cite{Becattini:2017gcx,Voloshin:2017kqp} illustrated in the left panel of Fig.~\ref{fig-argument} that for a non-central collision the collective elliptic flow generates vorticity $\omega^z$ and local angular momentum $L^z$ along the collision axis~$z$. According to this argument both $\omega^z$ and $L^z$ have a quadruple structure which means that when the plane transverse to the collision axis is divided into four quadrants, $\omega^z$ and  $L^z$ should be positive in the first and third quadrants, and negative in the second and fourth quadrants. The argument is not precise as it assumes that the collective flow is only along the $x$~axis and its gradient is along the $y$~axis. In other words, it takes into account only the first term in Eq.~(\ref{Lz-result}). In a physically realistic system the collective flow has both $x$ and $y$~components and the second term in Eq.~(\ref{Lz-result})  contributes with opposite sign, so it is not clear which term  will dominate in a given quadrant. It might happen that each quadrant splits into two domains so that $\omega^z$ and $L^z$ have not quadruple but octuple structure.  Such a situation is shown in the right panel of Fig.~\ref{fig-argument} where in addition to flow along the $x$~axis there is also flow along the $y$~axis. 

To simplify the notation in subsequent sections we replace the vector $\vec{R}$ which denotes position with respect to the center of the interaction region by $\vec{r}=(x,y)$. We also do not write explicitly the dependence of functions on the proper time and spatial rapidity. In all contour plots the axes show the transverse plane with distances given in femtometer (fm). 

The procedure we use to estimate errors in this section is slightly different from that of Sec.~\ref{Ly-sec}. The reason is that we calculate $L^z$ locally, in $\sim 1600$ small regions distributed across the transverse plane, which means the numerical calculation is more time consuming than the calculation of global $L^y$ by orders of magnitude. For this reason we do not use 51 values of $\bar\delta$ but instead use only three: calculations are done with $\bar\delta=0.9$ and error is estimated by comparison with the results obtained from $\bar\delta=0.8$ and $\bar\delta=1.0$.

%=========================================================================
\subsection{Vorticity }
\label{sec-vort}
%=========================================================================

In this section we define the glasma vorticity. To do this we must first decide how to define the velocity of the glasma. In our earlier studies of collective phenomena in  glasma \cite{Carrington:2021qvi,Carrington:2023nty,Carrington:2024utf} we defined 
the fluid velocity in the transverse plane as $\vec V(\vec r) \equiv \big(T^{0x}(\vec r),T^{0y}(\vec r)\big)/T^{00}(\vec r)$. Since the Poynting vector $\vec{P} \equiv (T^{0x}, T^{0y})$ is the energy flux and $T^{00}$ is the energy density, the ratio $\vec V \equiv \vec{P}/T^{00}$ defines the velocity in a natural way. Both $\vec{P}$ and $T^{00}$ come from a microscopic field theory applicable to any system independently of whether that system is in equilibrium or not. These are obvious advantages of our definition. However, there are also two evident deficiencies: $\vec V$ is not Lorentz covariant and there is no conserved quantity associated with it. 

Another alternative is to attempt to calculate a hydrodynamic four-velocity from the Landau-Lifshitz condition which would give the flux of energy. This kind of hydrodynamic definition of velocity was used in the context of glasma in \cite{Fries:2017ina,McDonald:2023qwc} and in our work \cite{Carrington:2024utf} where the fluid-like behavior of glasma was studied in detail. The hydrodynamic four-velocity $u^\mu(\vec r)$ can be defined to be the eigenvector of the energy-momentum tensor associated with the energy-density eigenvalue. It is important to remember that if the system is not in local thermal equilibrium, this eigenvector, which we call below a ``proxy hydrodynamic velocity,'' will not necessarily correspond physically to the fluid velocity. To test the reliability of the procedure we can use the proxy hydrodynamic velocity to construct a hydrodynamic EMT. One can either assume that the EMT has the ideal form, or one can use an anisotropic hydrodynamic form \cite{Florkowski:2010cf,Martinez:2010sc} which allows for a negative longitudinal pressure. 
A comparison of the glasma EMT and a hydrodynamic reconstruction of it can be used to evaluate the extent to which the glasma can actually be described as a fluid, and most importantly for our purposes, whether or not the proxy hydrodynamic velocity really has the physical meaning of a fluid velocity. 
The results show that while some components of the two EMT's are very similar, others are quite different. 
The $zz$ component of the glasma energy-momentum tensor, which is the longitudinal pressure, is negative and thus is qualitatively different than that of ideal hydrodynamics. 
Anisotropic hydrodynamics provides a much better description of the longitudinal pressure, but other components of the glasma EMT are still poorly described.  
The failure of hydrodynamics to reproduce the glasma EMT means  that the proxy hydrodynamic velocity does not posses all of the properties of the true hydrodynamic velocity. In this sense, the meaning of this flow vector is physically ambiguous. This is exactly as expected since hydrodynamics is valid only for systems in local thermal equilibrium, whereas the glasma is far from equilibrium. 

We will define the glasma vorticity along the $z$-axis as
\be
\label{vort-def}
\omega^z(\vec r) \equiv \big(\vec\nabla\times \vec V(\vec r)\big)^z  
= -\left(\frac{\partial V^x(\vec r)}{\partial y}-\frac{\partial V^y(\vec r)}{\partial x}\right) \,.
\ee 
We note that a factor 1/2 is often included on the right side of the definition (\ref{vort-def}).
We can compare this definition with the definition usually used in relativistic hydrodynamics, see the reviews \cite{Florkowski:2018fap,Becattini:2024uha}, which is 
\be
\label{thermal-vort}
\omega_{\rm th}^z(\vec r) \equiv \big(\vec\nabla\times \vec\beta(\vec r)\big)^z \,,
\ee
where
$\vec\beta(\vec r)$ is the spatial part of the four-vector $\beta^\mu(\vec r) \equiv u^\mu(\vec r)/T(\vec r)$ and $T(\vec r)$ is the position dependent temperature in the fluid local rest frame. If the temperature is constant then the two definitions of  vorticity differ only by a constant factor $1/T$. For an out of equilibrium system however the temperature is not even a well defined quantity. Hydrodynamic calculations have found that behaviour of the vorticity is strongly dependent on the spatial dependence of the temperature field, and that even the sign of the vorticity can change if the temperature is strongly nonuniform.  We comment that since the hydrodynamic approach assumes that thermodynamic variables like the temperature depend only weakly on spatial variables, the fact that a spatially dependent temperature can change the sign of the vorticity is a clear sign that a hydrodynamic approach may not be valid. 

The most direct way to understand the differences in the two definitions of vorticity is to compare the results obtained from them. The main result of our work \cite{Carrington:2024utf} is that the glasma exhibits a suprising degree of fluid-like behavior.  This happens because the glasma energy-momentum tensor satisfies the equations of universal flow \cite{Vredevoogd:2008id} that we discuss and use in Sec.~\ref{sec-compare}. We therefore expect that the two definitions of vorticity should give similar results. It is straightforward to use the definition (\ref{vort-def}) to calculate the vorticity from our glasma EMT. To calculate thermal vorticity (\ref{thermal-vort}) we must resolve two problems. The first is that we need a way to define the temperature of the far from equilibrium glasma. To obtain a crude estimate we can use the temperature of an equilibrated quark-gluon plasma whose energy density is the same as the energy density of the glasma, as described in Sec.~IVA of our paper \cite{Carrington:2022bnv}. The second difficulty is technical in nature. The problem is that the proxy hydrodynamic velocity is not smooth because the numerical uncertainty in the energy-momentum tensor is compounded in the velocity field through uncertainties introduced in the calculation of eigenvectors at every space-time point, and these errors further increase when we calculate numerical derivatives. A comparison of the two calculations shows that the thermal vorticity (\ref{thermal-vort}) has the same basic shape as the vorticity calculated from (\ref{vort-def}) (see Sec.~\ref{sec-vorticity}), but the resulting vorticity field is much too numerically unstable to allow for the accurate Fourier analysis we perform in Secs.~\ref{sec-fourier} and \ref{sec-vorticity}. This happens because the Fourier coefficients are sensitive to the detailed structure of the field.

We conclude that for practical reasons it is preferable to use the vorticity defined in (\ref{vort-def}) with the velocity $\vec V$ obtained directly from the glasma energy-momentum tensor. We note however that, due to the issues discussed above, the interpretation of glasma vorticity should be considered carefully.

%=========================================================================
\subsection{Local angular momentum}
\label{sec-Lz}
%=========================================================================

We calculate the $z$~component of the local angular momentum about a point $\vec r_0 = r_0 \big(\cos(\phi_0),\sin(\phi_0)\big)$ from the formula (\ref{Lz-result}) with $\vec R \equiv \vec r = \vec \rho -\vec r_0$ and the integration taken over a small area $d^2\rho$ around the point $\vec r_0$ in the transverse plane. The integration is done over a small square region and we use $\Delta$ for the size of the box in each dimension. The dependence of the result on the value of $\Delta$ is an important and subtle issue that is discussed in detail in Sec.~\ref{sec-compare}. We note that we actually compute not the angular momentum $L^z(\vec r_0)$ but the angular momentum per unit spatial rapidity $dL^z(\vec r_0)/d\eta$ as defined by Eq.~(\ref{Lz-result}). For simplicity however we will call this the angular momentum and write $L^z(\vec r_0)$. 

Vorticity and local angular momentum are expected to be closely related to each other. Physically the connection is easy to understand for the case of a rotating rigid body and it is commonly expected that the two quantities are also related to each other in more general situations. If we calculate $L^z(\vec r_0)$ from (\ref{Lz-result}) by performing a two variable Taylor expansion around the point $\vec r_0$ and integrating over a small box centered on the point $\vec r_0$ then the first term in the series that is not trivially zero has the form
\be
\label{leading-L}
L^z(\vec r_0) =  - \frac{\tau \Delta^4}{12} \left(\frac{\partial T^{x0}(\vec r)}{\partial y}-\frac{\partial T^{y0}(\vec r)}{\partial x}\right) \bigg|_{\vec r = \vec r_0}\,.
\ee 

A comparison of Eqs.~(\ref{vort-def}) and  (\ref{leading-L}) seems to indicate that the difference between vorticity and local angular momentum is analogous to the difference between the vorticity and thermal vorticity discussed above. It appears that $\omega^z(\vec r)$ and $L^z(\vec r_0)$ will behave similarly if the gradients of the energy density $T^{00}(\vec r)$ are negligible. This assumption seems reasonable and is likely to be well satisfied in systems that are not too far from equilibrium. In the glasma it is certainly true close to the center of the transverse plane but at the edges of the interaction region, where $\omega^z(\vec r)$ and $L^z(\vec r_0)$ reach their maximal values, the gradients of the energy density are not expected to be small. In addition, we know that there are constraints on the glasma energy-momentum tensor (for example, it is traceless and divergenceless) and these conditions might lead to significant cancellations between the two terms in (\ref{leading-L}) and make the apparent similarity of the expressions in  Eqs.~(\ref{vort-def}) and  (\ref{leading-L}) irrelevant. This is in fact what happens as we will explain in detail in Sec.~\ref{sec-compare}. 

%=========================================================================
\subsection{Fourier decomposition}
\label{sec-fourier}
%=========================================================================

To better understand the relationship between local angular momentum and vorticity, and the information about the system that they provide, we perform a Fourier decomposition of the velocity field in case of vorticity and of the Poynting vector field in case of local angular momentum. 

First we discuss vorticity. We write the velocity field in the transverse plane in polar coordinates as
\be
\label{velocity-field}
\begin{aligned}
V^x(\vec r) &=& \sum_{n=0}^\infty u^x_n(r) \cos(\phi) \cos(n\phi) ,
\\ 
V^y(\vec r) &=& \sum_{n=0}^\infty u^y_n(r) \sin(\phi)\cos(n\phi)  ,
\end{aligned}
\ee
where $\vec r = (r,\phi)$. Since we consider collisions of identical nuclei the velocity field has the symmetry $\vec{V}(\vec{r}) = - \vec{V}(-\vec{r})$ and consequently the summations are only over even $n$ in Eq.~(\ref{velocity-field}). The Fourier components $u_n^x(r)$ and $u_n^y(r)$ can be obtained from the velocity field $\vec{V}(\vec r)$ as
\be
\label{Vn-extract}
\begin{aligned}
u_0^x(r) &= \int_0^{2\pi} \frac{d\phi}{2\pi}\, V^x(\vec r)\,\frac{1}{\cos(\phi)} \,,
~~~~~~~~~
u_0^y(r) = \int_0^{2\pi} \frac{d\phi}{2\pi}\, V^y(\vec r)\,\frac{1}{\sin(\phi)} \,,
\\[4mm]
u_{n>0}^x(r) &= \int_0^{2\pi} \frac{d\phi}{\pi}\, V^x(\vec r)\,\frac{\cos(n\phi)}{\cos(\phi)} \,,
~~~~~
u_{n>0}^y(r) = \int_0^{2\pi} \frac{d\phi}{\pi}\, V^y(\vec r)\,\frac{\cos(n\phi)}{\sin(\phi)} \,.
\end{aligned}
\ee

When $u_n^x(r) = u_n^y(r) = u_n(r)$ the velocity $\vec V(\vec r)$ is along the vector $\vec r$ but its magnitude depends on both $r$ and $\phi$. Such a flow is radial  $(\vec V(\vec r) \parallel \vec r)$ but not cylindrically symmetric ($|\vec V(\vec r)|$ is not independent of $\phi$ unless $u_{n\ge 2}=0$). In this case the $z$~component of the vorticity equals 
\be
\label{vort-radial-flow}
\omega^z(\vec r) =  \frac{1}{r} \sum_{n=0}^\infty n \, u_n(r) \, \sin(n\phi) .
\ee
The $n$-th Fourier component of the velocity field generates an $n$-th Fourier component of the vorticity. Cylindrically symmetric flow ($u_0\ne 0$ and $u_{n>0}=0$) gives, as expected, vanishing $\omega^z$. A nonzero $u_2$ coefficient, or a quadrupole contribution to the velocity flow, produces a quadrupole structure for $\omega^z$, and $u_4$ generates an octupole contribution to $\omega^z$, etc.

The glasma velocity field cannot be accurately decomposed into Fourier harmonics using the formulas (\ref{velocity-field}) if it is assumed that $u_n^x(r) = u_n^y(r)$. This happens because the transverse Poynting vector $\vec P(\vec r)$ is not always parallel to $\vec r$. To take into account a contribution to the velocity field that is not parallel to $\vec r$ we keep $u_n^x(r) \not= u_n^y(r)$ and express the coefficients as
\be
 u_n^x(r) = u_n(r)+\delta_n(r),
 ~~~~~~~~~~~~~~~~~~
 u_n^y(r) = u_n(r)-\delta_n(r).  
\label{vdel-def}
 \ee
The vorticity then takes the form
\be
\label{vort-flow}
\omega^z(\vec r) = \sum_{n=0}^\infty \Big[
\frac{n}{r}\,\sin(n\phi) \big(u_n(r) + \cos(2\phi)\delta_n(r)\big)
-\cos(n\phi)\sin(2\phi) \Big(\frac{\delta_n(r)}{r} -  \frac{d\delta_n(r)}{dr}\Big) \Big] .
\ee
Keeping only the terms $n=0, 2, 4$ in Eq.~(\ref{vort-flow}) gives
\ba
\nn
&&\omega^z(\vec r)
= \Big(\frac{1}{r}\, \delta_0(r) - \frac{d \delta_0(r)}{d r}
+ \frac{2}{r} \, u_2 (r) + \frac{3}{2r}\, \delta_4(r) 
+ \frac{1}{2} \frac{d\delta_4(r)}{d r} 
\Big) \sin(2\phi) 
\\[2mm] \label{vort-flow-n=024}
&&~~~~~~~~
+ \, \frac{1}{2} \Big( \frac{3\delta_2(r)}{r} - \frac{d\delta_2(r)}{d r} 
+ \frac{8}{r} u_4(r) \Big) \sin(4\phi) 
+  \frac{1}{2} \Big(\frac{5}{r} \delta_4(r) 
-\frac{d \delta_4(r)}{d r} \Big) \sin(6\phi).\label{model-vort}
\ea
In contrast to the formula (\ref{vort-radial-flow}) the quadrupole contribution is generated not only by $u_2(r)$ but also by $\delta_0(r)$ and $\delta_4(r)$ and their derivatives. Similarly the octupole contribution to $\omega^z(\vec r)$ occurs not only due to $u_4(r)$ but also due to $\delta_2(r)$ and its derivative. This shows that the non-radial contributions to the velocity flow significantly complicate the structure of vorticity.

To perform the Fourier decomposition of the local angular momentum we proceed in exactly the same way but in this case we write the Poynting vector as a sum of harmonic components. In Eqs.~(\ref{velocity-field})-(\ref{vdel-def}) we make the replacements $(V^i(r,\phi),u_n(r),\delta_n(r)) \to (T^{i0}(r,\phi),p_n(r),\epsilon_n(r))$. We comment that although the Fourier coefficients $p_n(r)$ have the same physical meaning as the coefficients of the glasma collective flow $v_n$ studied in our earlier work \cite{Carrington:2021qvi,Carrington:2023nty,Carrington:2024utf}, they are differently defined.  The angular distribution of a collective flow is usually decomposed into Fourier components as 
\be
{\cal P}(\phi) = \frac{1}{2\pi} \Big(1 +  2\sum_{n=1}^\infty v_n \cos(n \phi ) \Big). 
\ee
These coefficients $v_n$ are related to our $p_n$ as $2 v_n = p_n/p_0$ for $n>0$.
The difference between our coefficients $u_n$ and $p_n$ is only the weighting function which is $\sqrt{\big(T^{0x}(\vec r)\big)^2 + \big(T^{0y}(\vec r)\big)^2}$ in the case of the $p_n$ coefficients, and $\sqrt{\big(T^{0x}(\vec r)\big)^2 + \big(T^{0y}(\vec r)\big)^2}/T^{00}(\vec r)$ in the case of the $u_n$.

We repeat the calculations that led from Eq.~(\ref{Lz-result}) to (\ref{leading-L}) and take into account not only terms of order $\Delta^4$ but also $\Delta^6$. This requires expanding the Poynting vector up to the third order and gives for $n=0, 2, 4$ the result 
\bea
L^z(\vec r) &=& \tau\frac{\Delta^4}{12} \Big[l_2^{(4)}(r) \sin(2\phi) + l_4^{(4)}(r)  \sin(4\phi) 
+  l_6^{(4)}(r) \sin(6\phi)\Big] \nn 
\\ [2mm] 
\label{modelLz}
&+& \tau\frac{\Delta^6}{120} \Big[l^{(6)}_2(r)\sin(2\phi)+l^{(6)}_4(r)\sin(4\phi)+l^{(6)}_6(r)\sin(6\phi)\Big]
\eea
with
\ba
\label{modelLz2}
\begin{aligned}
l_2^{(4)}(r) &= \frac{1}{r}\,\epsilon_2(r) - \epsilon_2'(r)
+ \frac{2}{r} \, p_2 (r) + \frac{3}{2r}\, \epsilon_4(r) 
+ \frac{1}{2} \epsilon_4'(r)  ,
\\[3mm]
l_4^{(4)}(r) &= \frac{1}{2} \Big( \frac{3}{r}\,\epsilon_2(r) - \epsilon_2'(r) 
+ \frac{8}{r}\, p_4(r) \Big)  ,
\\[3mm]
l_6^{(4)}(r) &= \frac{1}{2} \Big(\frac{5}{r} \epsilon_4(r) - \epsilon_4'(r) \Big)  ,
\\[3mm]
l^{(6)}_2(r) &= -\frac{29}{16 r^3}\,  p_2(r) - \frac{25}{48 r^2}\, p_2'(r) 
+ \frac{7}{12 r}\, p_2''(r) -\frac{1}{48} p_2'''(r)
\\[1mm]
&~~~\, -\frac{1}{r^3}\,\epsilon _0(r) - \frac{1}{r^3}\,\epsilon_4(r) 
+\frac{1}{r^2}\epsilon_0'(r) - \frac{1}{3 r^2}\,\epsilon_4'(r) 
+\frac{5}{6 r}\,\epsilon_4''(r)  - \frac{1}{3} \epsilon_0'''(r)
+\frac{1}{6}\epsilon_4'''(r) ,
\\[3mm]
l^{(6)}_4(r) &= \frac{5}{8 r^3}\, p_0(r) - \frac{35}{2 r^3}\, p_4(r) 
- \frac{5}{8 r^2}\,  p_0'(r) - \frac{7}{6 r^2}\,p_4'(r) + \frac{1}{4 r}\,p_0''(r)
+ \frac{7}{6 r}\,p_4''(r) - \frac{1}{24}\,p_0'''(r)  
\\[1mm]
&~~~\,- \frac{25}{4 r^3}\,\epsilon _2(r) + \frac{19}{12r^2}\,\epsilon _2'(r)
+\frac{5}{12 r}\,\epsilon _2''(r) -\frac{1}{6}\,\epsilon _2'''(r) ,
\\[3mm]
l^{(6)}_6(r) &= \frac{35}{16 r^3}\, p_2(r) - \frac{19}{16 r^2}\, p_2'(r)
+\frac{1}{4r}\,p_2''(r) - \frac{1}{48}\,p_2'''(r) 
\\[1mm]
&~~~\,- \frac{70}{3 r^3}\,\epsilon _4(r) +\frac{10}{3 r^2}\,\epsilon _4'(r) 
+ \frac{5}{6 r}\,\epsilon_4''(r) - \frac{1}{6} \epsilon_4'''(r)  ,
\end{aligned}
\ea
where we use primes to denote derivatives with respect to $r$. We note that the structure of the Fourier decomposition  of the local angular momentum is more complicated than that of vorticity as not only $p_n(r),~\epsilon_n(r)$ and $\epsilon'_n(r)$ but also $p'_n(r)$ contribute to the Fourier components of $L^z(\vec r)$. Note however that the first line in Eq.~(\ref{modelLz}) with the coefficients given in (\ref{modelLz2}) matches exactly the result in Eq.~(\ref{vort-flow-n=024}) with $u_n(r)$ replaced by $p_n(r)$ and $\delta_n(r)$ by $\epsilon_n(r)$. This is a simple consequence of the mathematical relationship between the two quantities (see Eqs.~(\ref{vort-def}) and (\ref{leading-L})). It leads to the expectation that $L^z/\Delta^4$ should behave very similarly to vorticity in the limit that a very small region is used to do the integration. In fact this is not the case, as will be explained in Sec.~\ref{sec-compare}.

%=========================================================================
\subsection{Glasma vorticity}
\label{sec-vorticity}
%=========================================================================

\begin{figure}[t]
\centering
\includegraphics[width=9.5cm]{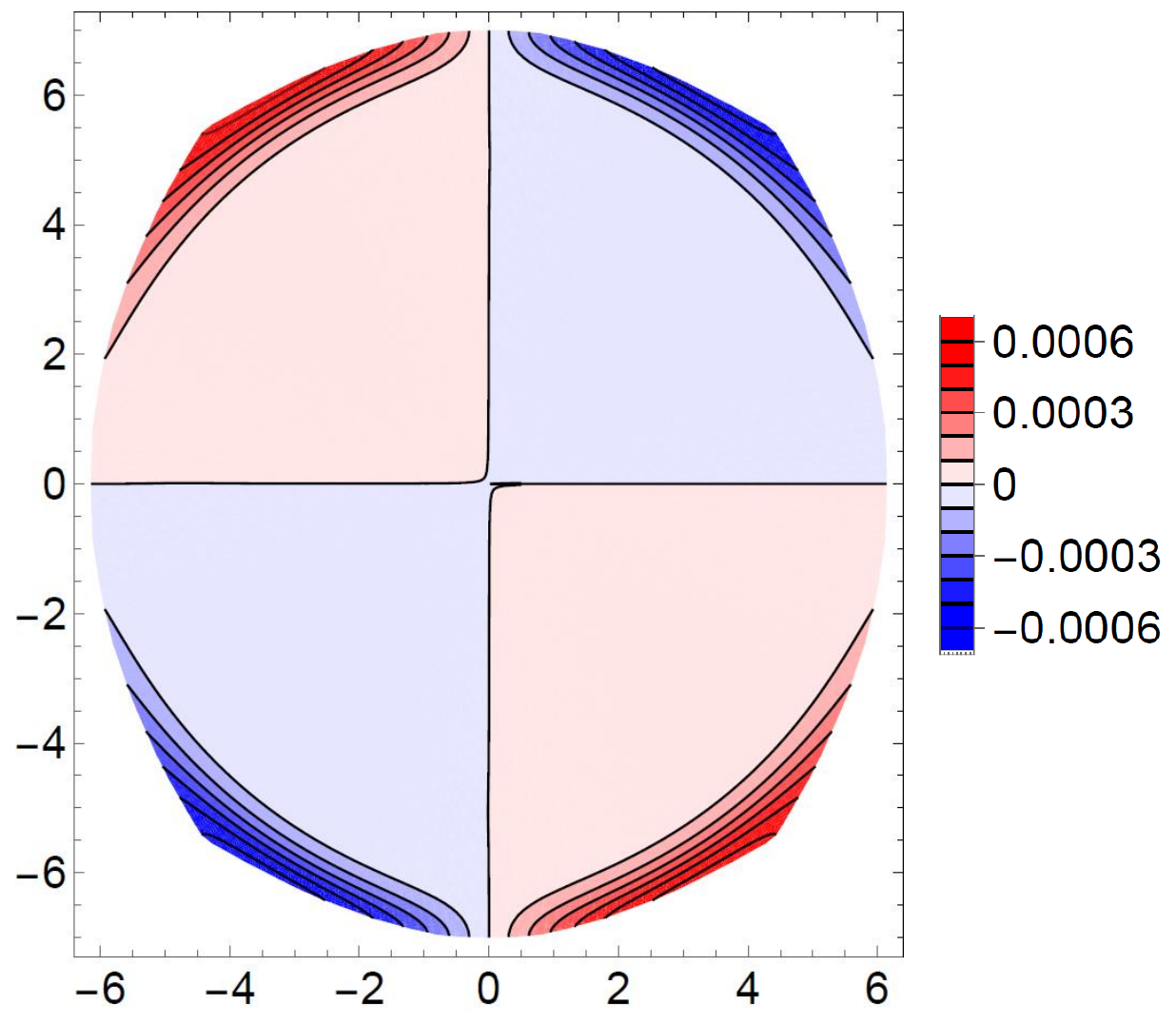}
\caption{Vorticity in the transverse plane expressed in ${\rm fm}^{-1}$ for $b=2$~fm. 
\label{fig-vort}}
\end{figure}

We calculate the vorticity of the glasma using Eq.~(\ref{vort-def}). The glasma vorticity for collisions with impact parameter $b=2$~fm is shown in Fig.~\ref{fig-vort}. The plot includes all points in the transverse plane for which $\delta(r_x,r_y) \le 1.0$, which is the region where the gradient expansion can be trusted. One sees a clear quadrupole structure, which agrees with the pattern of the $\Lambda$ polarization along the beam direction observed by the STAR and ALICE Collaborations, see \cite{STAR:2019erd,STAR:2023eck} and \cite{ALICE:2021pzu}, respectively, but the sign of the glasma vorticity is the opposite of what is seen experimentally. It is also opposite to that of the mechanism illustrated in the left panel of Fig.~\ref{fig-argument}. Interestingly, the sign of the glasma vorticity agrees with the sign of the thermal vorticity obtained in hydrodynamical models of relativistic heavy-ion collisions, see e.g. Fig.~6 of \cite{Becattini:2021iol}. We also note that when thermal vorticity is calculated from the glasma energy-momentum tensor, the result is qualitatively very similar to Fig.~\ref{fig-vort}. The calculation is described in Sec.~\ref{sec-vort} where we explain why the numerical errors are extremely difficult to control.

\begin{figure}[t]
\centering
\includegraphics[width=9cm]{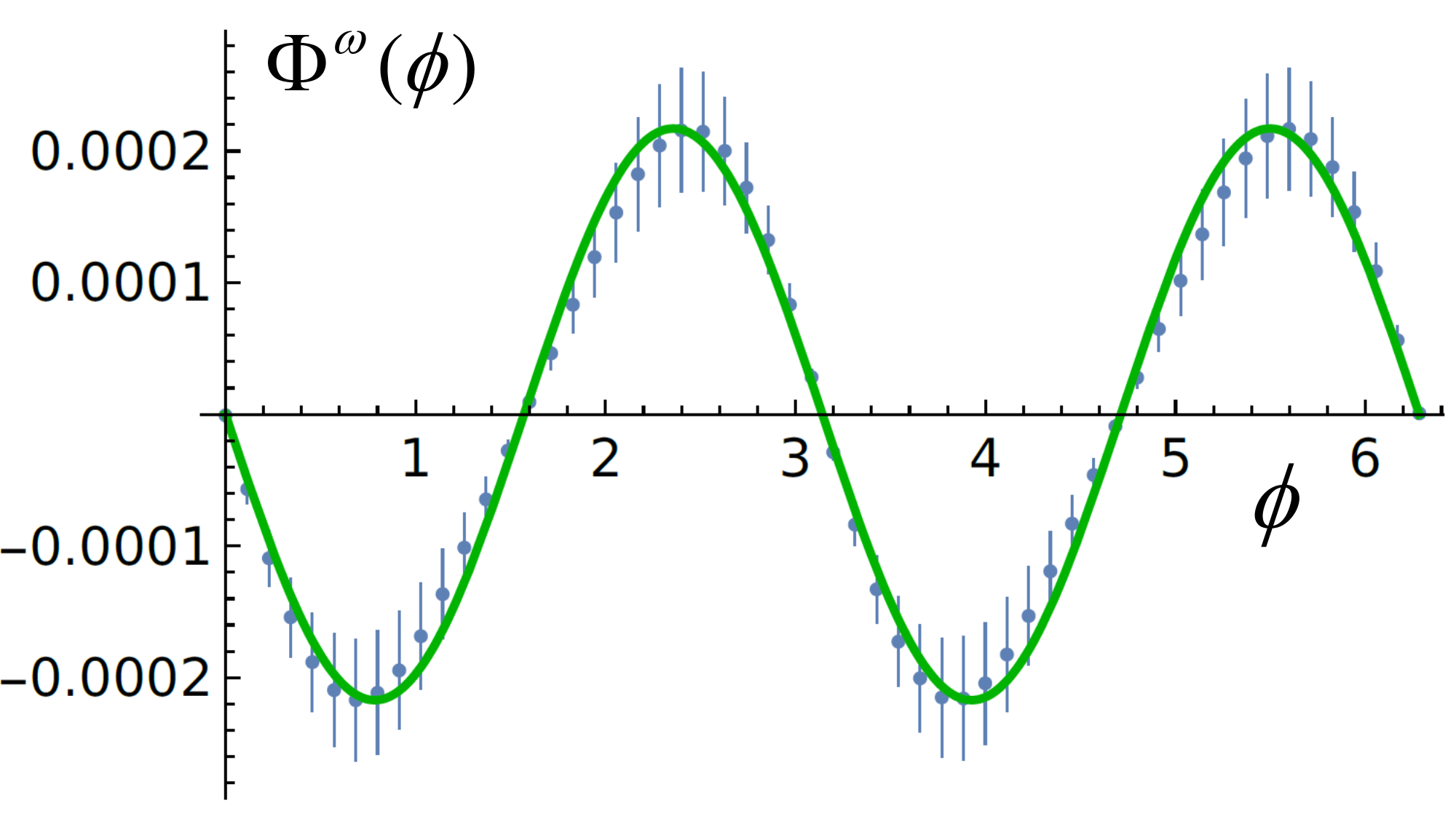}
\vspace{-1mm}
\caption{Angular dependence of the glasma vorticity from Fig.~\ref{fig-vort} together with a green curve showing the function $\sin(2\phi)$ with amplitude set to match the data.}
\label{fig-vort-angular}
\end{figure}

To verify the quadrupole structure we calculate the vorticity integrated over the radial coordinate 
\be
\Phi^\omega(\phi) \equiv \int_0^{R_{\rm max}} dr \, r\,\omega^z(r,\phi) .
\ee
We use $R_{\rm max}$ equal to the maximum $r$ for which $\delta(r_x,r_y) \le 0.9$ for all values of $\phi$. The result corresponding to Fig.~\ref{fig-vort} is shown in Fig.~\ref{fig-vort-angular}. To obtain the error bar we calculate $\Phi^\omega(\phi)$ with $\bar\delta = 0.8$ and $\bar\delta = 1.0$ and calculate the average of the absolute value of the difference with the result obtained with $\bar\delta = 0.9$. Fig.~\ref{fig-vort-angular} clearly shows that the quadrupole contribution to the vorticity is dominant. To quantify the $n$-pole contributions we calculate
\be
\Phi^\omega_n \equiv \int_0^{2\pi} d\phi \, \sin(n\phi) \, \Phi^\omega(\phi)
\ee
for $n=2$ and $n=4$ which gives $\Phi^\omega_2/\Phi^\omega_4 \approx 9.5$.  

\begin{figure}[b]
\centering
\includegraphics[width=17cm]{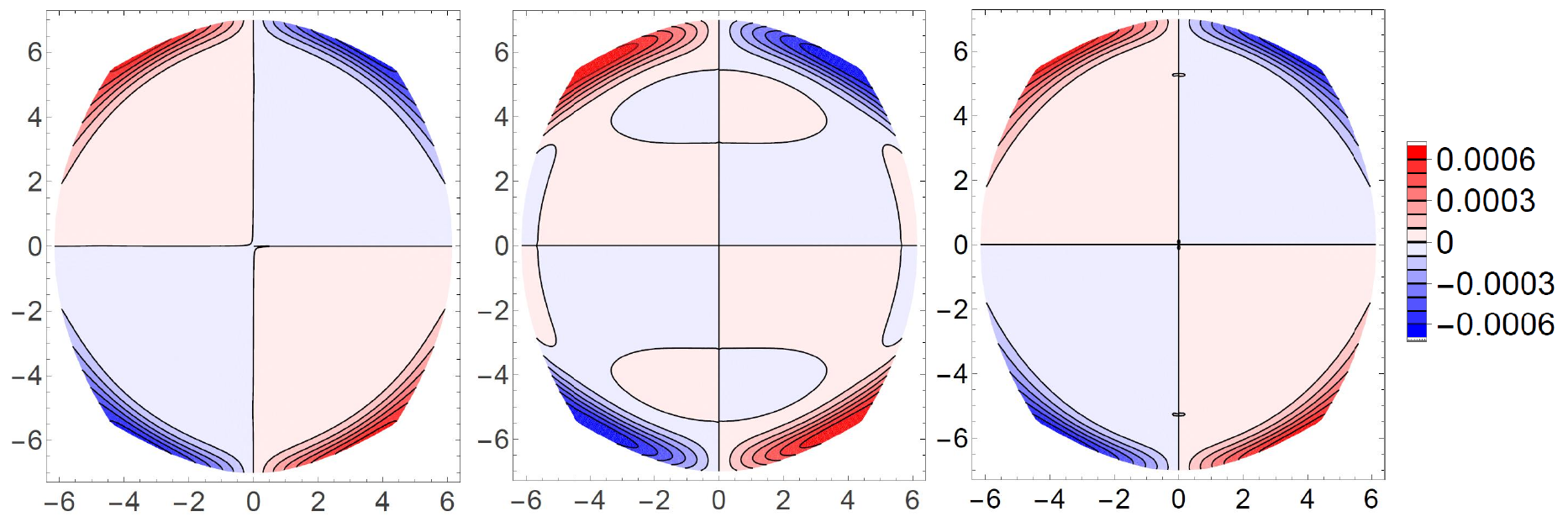}
\vspace{-2mm}
\caption{The vorticity from the glasma velocity field (left panel) and from the Fourier components $n=0,2$ (middle panel) and $n=0,2,4$ (right panel). }
\label{fig-vort-glasma-model}
\end{figure}

We can use the formula (\ref{vort-flow-n=024}) resulting from the Fourier decomposition to understand why glasma vorticity has the opposite sign from the experiment, and from the prediction of the simple physical argument described in Fig.~\ref{fig-argument}. First we verify that we can reproduce the results of our calculation using the Fourier coefficients. In the left panel of Fig.~\ref{fig-vort-glasma-model} we show the actual glasma vorticity (already shown in Fig.~\ref{fig-vort}), and the middle and right panels show the vorticity obtained with the $n=0,2$ and $n=0,2,4$ Fourier components, respectively. The figure demonstrates that the glasma vorticity is well reproduced with the first three Fourier harmonics. Since the right panel of Fig.~\ref{fig-vort-glasma-model} cannot be distinguished visually from the left panel, we calculate the average of the absolute value of the difference of the two sets of data, divided by the average of the absolute value of the sum. This gives 0.025. 

\begin{figure}[t]
\centering
\includegraphics[width=16cm]{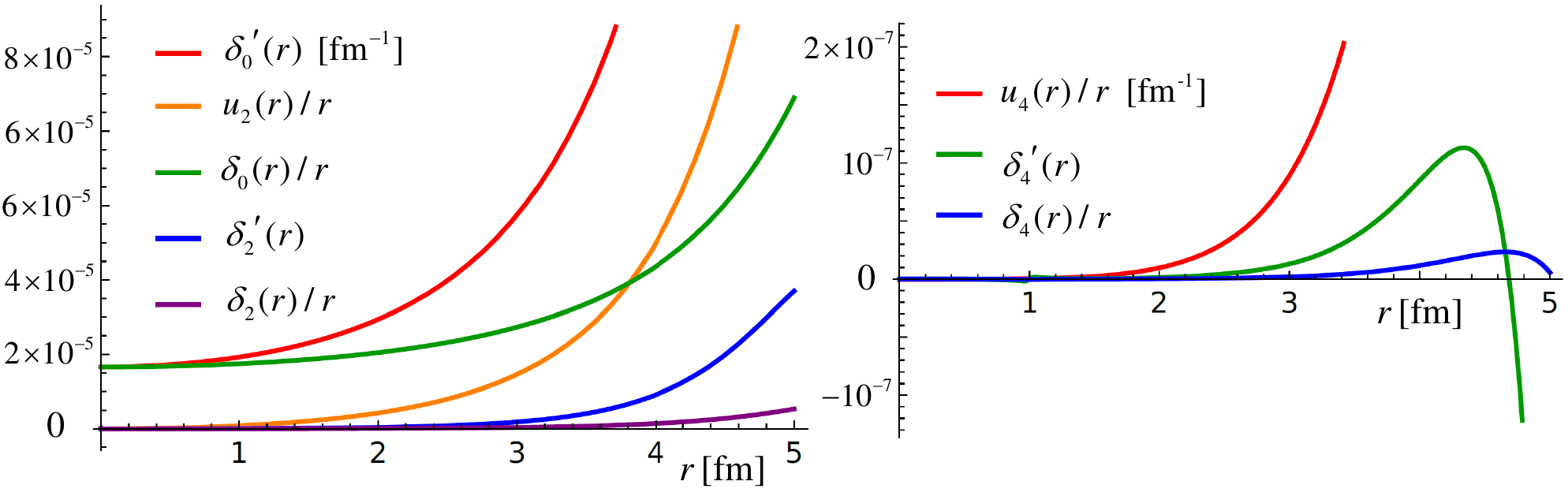}
\vspace{-2mm}
\caption{The Fourier coefficients of the glasma velocity field: $\delta_0(r)/r, ~\delta_0'(r), ~ u_2(r)/r,~\delta_2(r)/r, ~\delta_2'(r)$  (left panel) and $u_4(r)/r, ~\delta_4(r)/r, ~\delta_4'(r)$ (right panel).}
\label{plot-coefs}
\end{figure}

\begin{figure}[b]
\centering
\includegraphics[width=16cm]{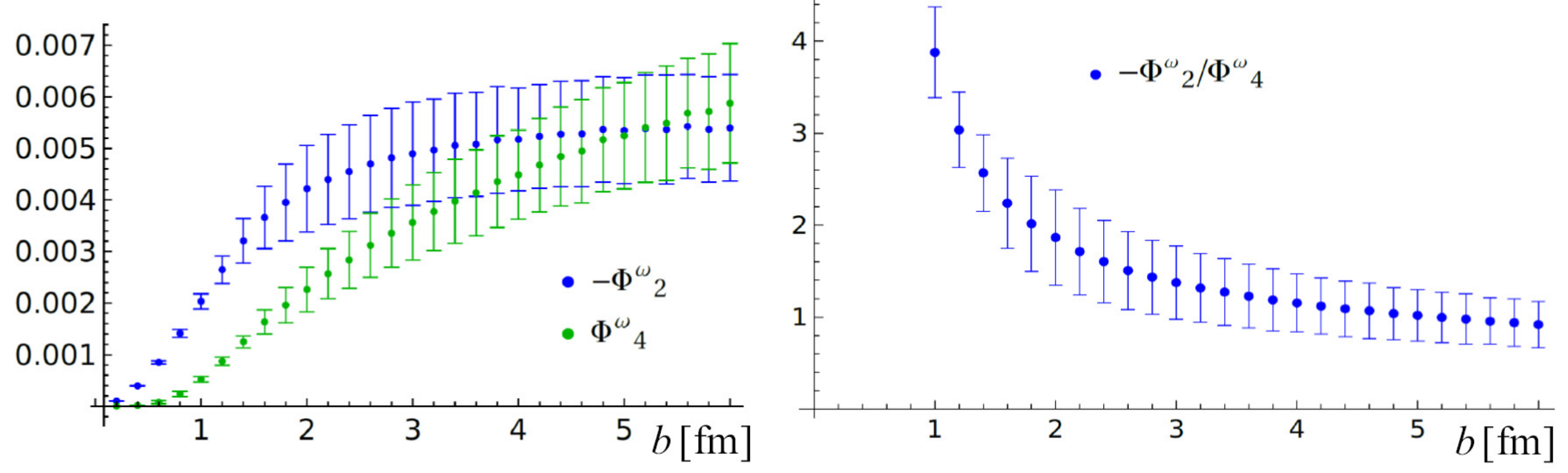}
\caption{The quantities $-\Phi^\omega_2$, $\Phi^\omega_4$ (left) and $-\Phi^\omega_2/\Phi^\omega_4$ (right) versus $b$.}
\label{plotW-bvals}
\end{figure} 

We can also identify the components of the Fourier decomposition that determine the pattern of the glasma vorticity. In Fig.~\ref{plot-coefs} we show the Fourier components $u_n(r)$ and $\delta_n(r)$ as functions of $r$. Since we observe a quadruple structure we know that the term in Eq.~(\ref{vort-flow-n=024}) proportional to $\sin(2\phi)$ dominates. Figure~\ref{plot-coefs} shows that the largest contribution comes from $-\delta_0'(r)$ which is responsible for the negative sign of the vorticity in the first quadrant. This means that the quadruple structure of the vorticity does not come from the elliptic flow represented by $u_2(r)$ but instead from $\delta_0(r)$ which grows with $r$. In other words, the sign of glasma vorticity is opposite to that generated by the elliptic flow because the vorticity is dominated by the non-radial contribution which is bigger in the $x$ than in the $y$~direction. 

The results presented so far in this section are for impact parameter $b = 2$~fm. The vorticity structure changes significantly with the impact parameter. At small $b$ the quadrupole contribution dominates but when $b$ increases the octupole contribution grows and becomes equal to the quadrupole term at $b \approx 5$~fm. This is demonstrated in Fig.~\ref{plotW-bvals} where $-\Phi^\omega_2$, $\Phi^\omega_4$ and the ratio $-\Phi^\omega_2/\Phi^\omega_4$ versus $b$ are shown. The results are obtained using $\delta=0.9\pm 0.1$.

%=========================================================================
\subsection{Glasma local angular momentum}
\label{sec-Lz}
%=========================================================================

As discussed in section \ref{sec-Lz}, the behaviour of the local angular momentum is expected to be similar to the vorticity. In this section we present our results for $L^z(\vec r)$ and show that contrary to expectation angular momentum is significantly different from vorticity. In Sec.~\ref{sec-compare} we explain why. 

\begin{figure}[t]
\begin{center}
\includegraphics[width=16cm]{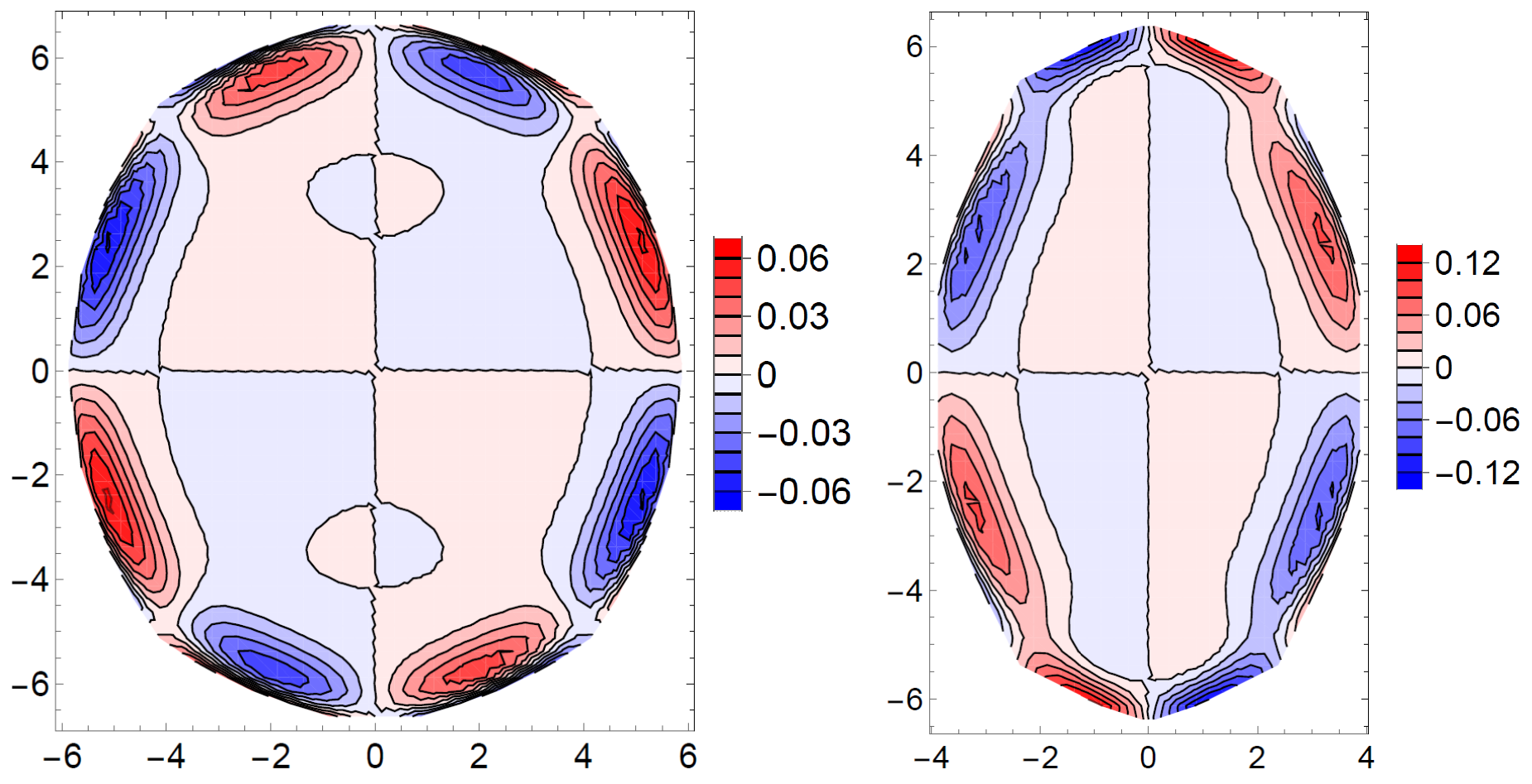}
\end{center}
\vspace{-7mm}
\caption{$L^z(\vec r)/\Delta^6$ in fm$^{-6}$ from collisions at $b=2$~fm (left) and $b=6$~fm (right). }
\label{plotcc} 
\end{figure}

The local angular momentum defined by Eq.~(\ref{Lz-result}) obviously depends on the transverse area over which the integral is performed. Our calculation of $L^z(\vec r)$ is done with a grid of 64$\times$64 points evenly distributed in the transverse plane and a square integration region of size $\Delta^2$ with $\Delta=0.2$~fm centered on each point. We include only boxes for which $\delta<1$ for all points in the box. As discussed in Sec.~\ref{Ly-sec}, the part of the transverse plane for which $\delta<1$ coincides well with the interaction region. In the next section we show that $L^z(\vec r)$ grows with the size of the area as $\Delta^6$ when $\Delta$ is sufficiently small. To make our results independent of $\Delta$ we divide $L^z(\vec r)$ by $\Delta^6$.

\begin{figure}[b]
\centering
\includegraphics[width=16cm]{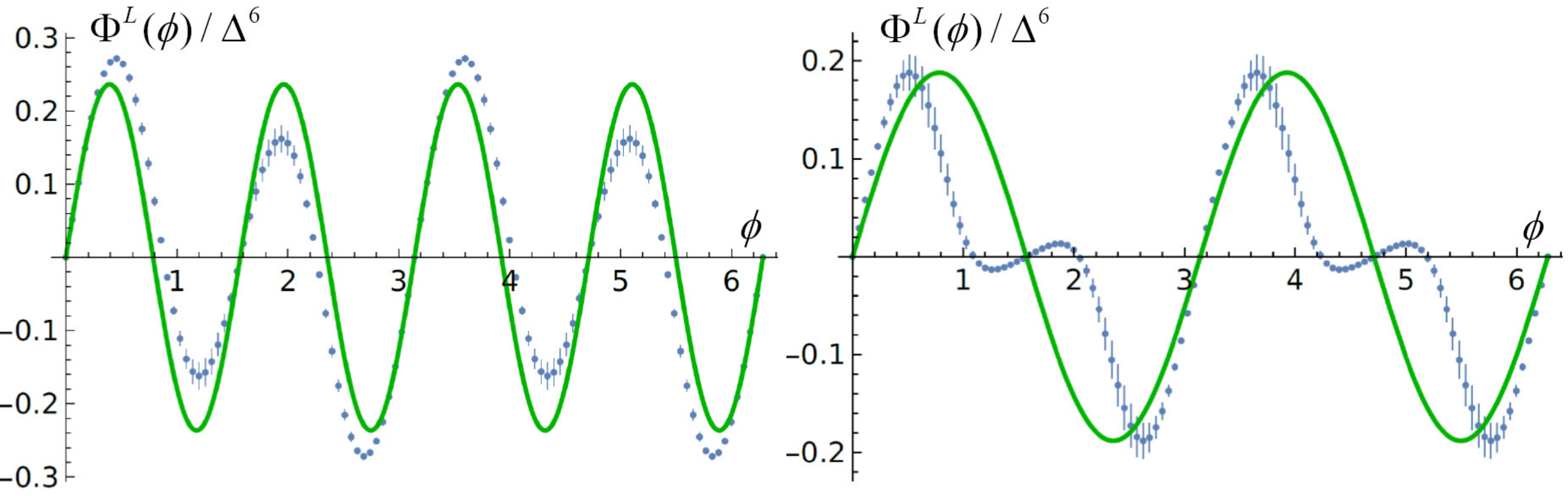}
\vspace{-1mm}
\caption{Angular dependence of $L^z(\vec r)/\Delta^6$ for $b=2$~fm (left) and $b=6$~fm (right).  The  green curves represent $\sin(4\phi)$ (left) and $\sin(2\phi)$ (right), with amplitudes set to match the data.}
\label{fig-Lz-angular}
\end{figure}

In Fig.~\ref{plotcc} we show $L^z(\vec r)/\Delta^6$ for collisions at $b=2$ fm (left panel) and $b=6$ fm (right panel). Comparing Figs.~\ref{plotcc} and \ref{fig-vort} one clearly sees that $L^z(\vec r)$ is qualitatively different from $\omega^z(\vec r)$. The left  panel of Fig.~\ref{plotcc} shows an octupole structure for $b=2$~fm but in the right panel one sees a combination of octupole and quadrupole patterns for $b=6$~fm. To quantify these observations we calculate
\be
\Phi^L(\phi) = \int_0^{R_{\rm max}} dr\,r\,L^z(r,\phi).
\ee
The upper limit of the integral is chosen as for the analogous vorticity calculation in Sec.~\ref{sec-vorticity}: $R_{\rm max}$ equals the maximum $r$ for which $\delta<0.9$ for all values of $\phi$. The function $\Phi^L(\phi)$ divided by $\Delta^6$ is shown in Fig.~\ref{fig-Lz-angular} for $b=2$~fm (left panel) and $b=6$~fm (right panel). The error is calculated using $\bar\delta=0.9\pm 0.1$. The octupole contribution to $L^z(\vec r)$ is clearly dominant for $b=2$~fm but for $b=6$~fm the superposition of quadrupole and octupole contributions is visible.

\begin{figure}[t]
\centering
\includegraphics[width=8cm]{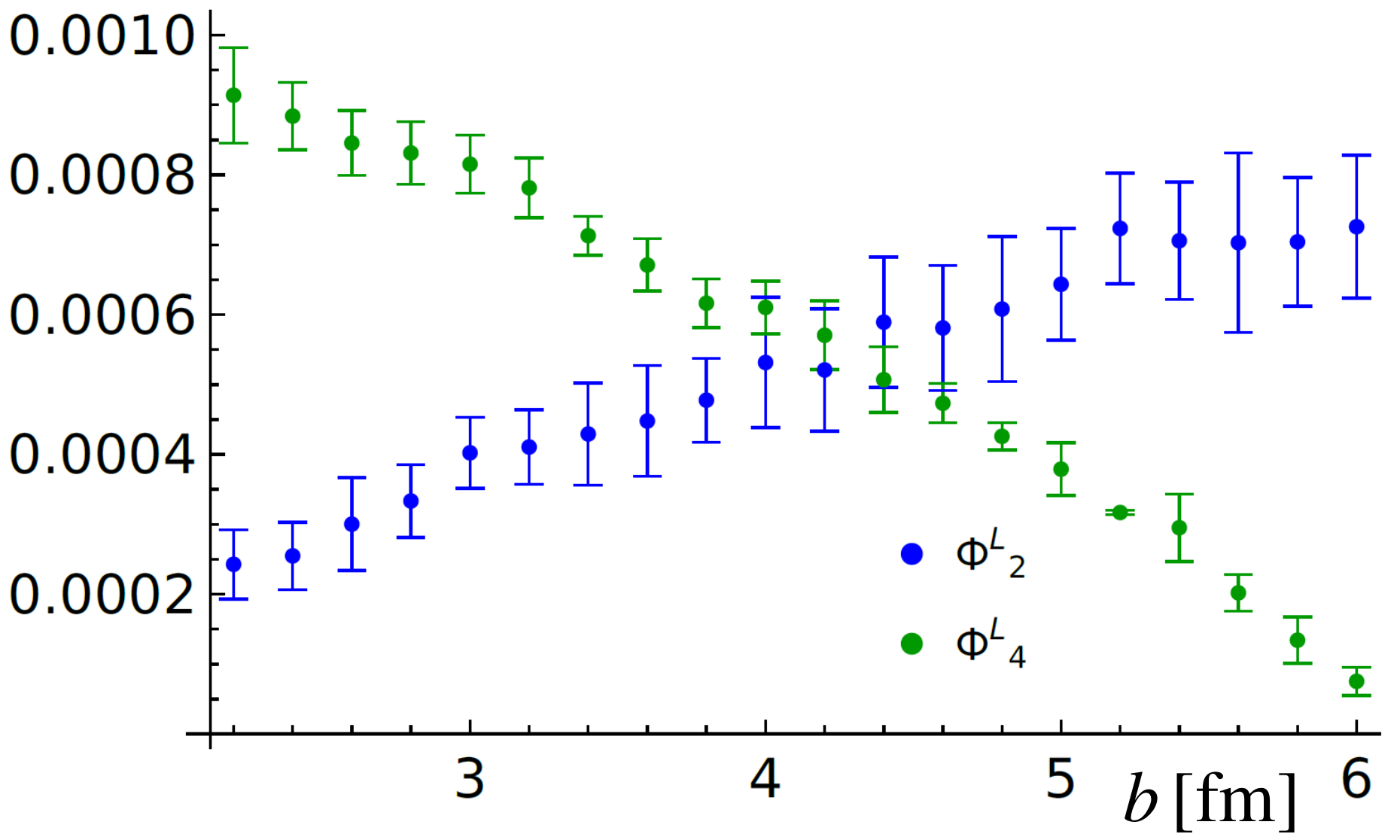}
\vspace{-3mm}
\caption{$\Phi^L_2$ and $\Phi^L_4$ versus $b$.}
\label{plotLz-bvals}
\end{figure}

To further quantify the $n$-pole contributions to local angular momentum we calculate
\be
\Phi^L_n \equiv \int_0^{2\pi} d\phi \, \sin(n\phi) \, \Phi^L (\phi),
\label{L-mom-def}
\ee
which is shown as a function of $b$ in Fig.~\ref{plotLz-bvals} for $n=2$ and $n=4$. The error is estimated by calculating $\Phi^L_n$ for $\bar\delta=0.8$ and $\bar\delta=1.0$. The octupole contribution decreases and the quadrupole term increases, as functions of impact parameter. It is important to emphasis that the sign of the quadrupole contribution is opposite to that of the vorticity and agrees with the sign of the experimentally observed $\Lambda$~polarization. These results reflect the fact that the naive argument illustrated in the left panel of Fig.~\ref{fig-argument} is correct at large impact parameter where the elliptic flow is strong enough that the quadrupole contribution dominates the vorticity. 

%=========================================================================
\subsection{Vorticity vs. local angular momentum}
\label{sec-compare}
%=========================================================================

As explained in the introduction, there are physical reasons to believe that the behaviour of vorticity and angular momentum are similar. Furthermore, the equations (\ref{vort-def}), (\ref{leading-L}), and the Fourier decomposed expressions (\ref{model-vort}), (\ref{modelLz}) and (\ref{modelLz2}), seem to indicate that this expectation is reflected in the mathematical structure of $\omega^z(\vec r)$ and $L^z(\vec r)$. The results in Secs.~\ref{sec-vorticity} and \ref{sec-Lz} however show that vorticity and local local angular momentum of glasma are qualitatively different. In this section we explain why. 

\begin{figure}[t]
\centering
\includegraphics[width=16cm]{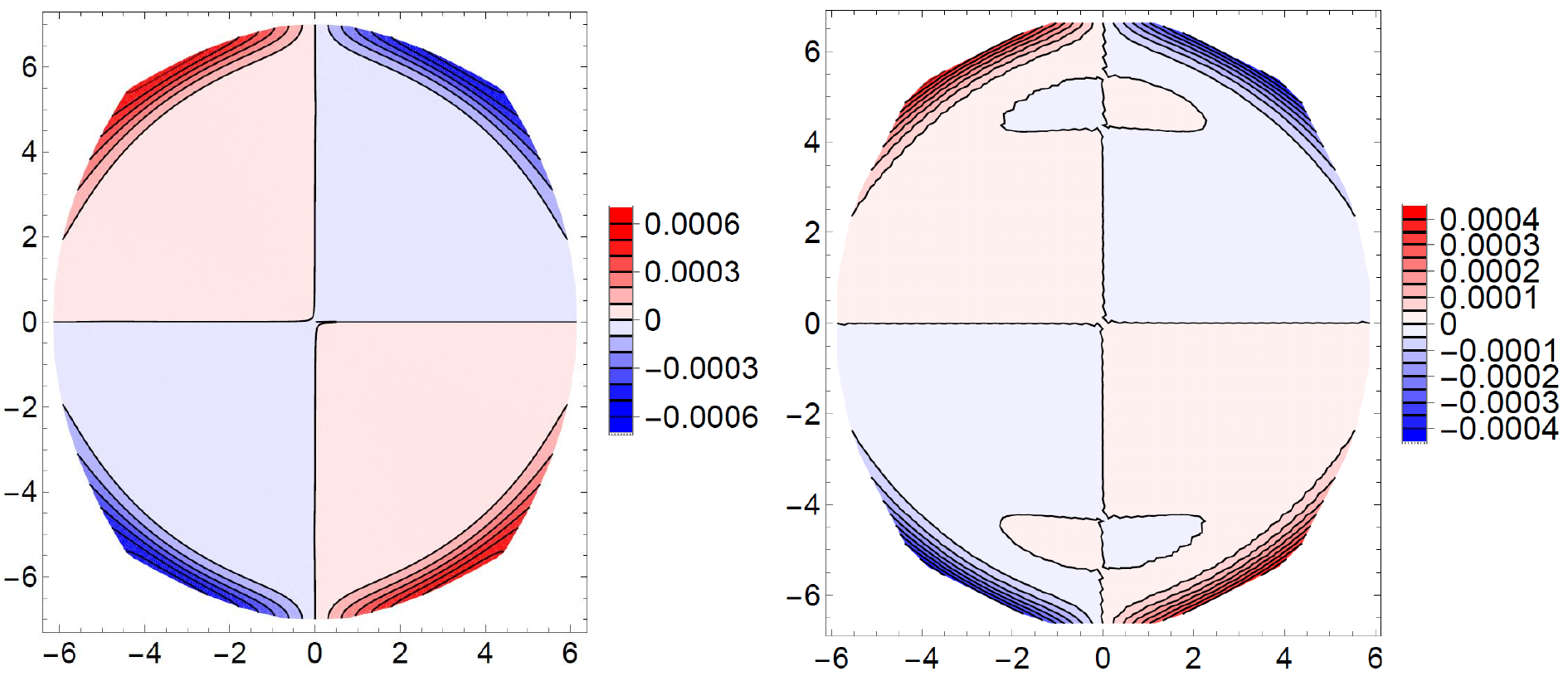}
\caption{The left panel shows  vorticity expressed in ${\rm fm}^{-1}$ (shown previously in Fig.~\ref{fig-vort}) and the right panel is $L^z_{\rm mod}(\vec r)$ scaled so that the average of the absolute value matches the vorticity.
\label{plotbb}}
\end{figure}

To investigate the simplest possible reason for the difference between $\omega^z(\vec r)$ and $L^z(\vec r)$ we calculate  what we call `modified' local angular momentum, and denote $L^z_{\rm mod}(\vec r)$, which is defined as in Eq.~(\ref{Lz-result}) but with the Poynting vector $\vec P(\vec r) \equiv \big(T^{0x}(\vec r),T^{0y}(\vec r)\big)$ replaced by the velocity field $\vec V(\vec r) \equiv \big(T^{0x}(\vec r),T^{0y}(\vec r) \big)/T^{00}(\vec r)$. The calculation is done in the same way as local angular momentum, as described in Sec.~\ref{sec-Lz}. The glasma vorticity and modified local angular momentum both for $b=2$~fm are compared in Fig.~\ref{plotbb}. Since local angular momentum depends on the size of the box used to calculate it and vorticity has dimension inverse length, we scale the data obtained for $L^z_{\rm mod}(\vec r) $ by the ratio of the average of the absolute values of the two data sets. Fig.~\ref{plotbb} confirms that $\omega^z(\vec r)$ and $L^z_{\rm mod}(\vec r)$ are qualitatively similar. We have also checked that $L^z_{\rm mod}(\vec r)/\Delta^4$ is approximately independent of $\Delta$, as expected. 

\begin{figure}[t]
\centering
\includegraphics[width=16.5cm]{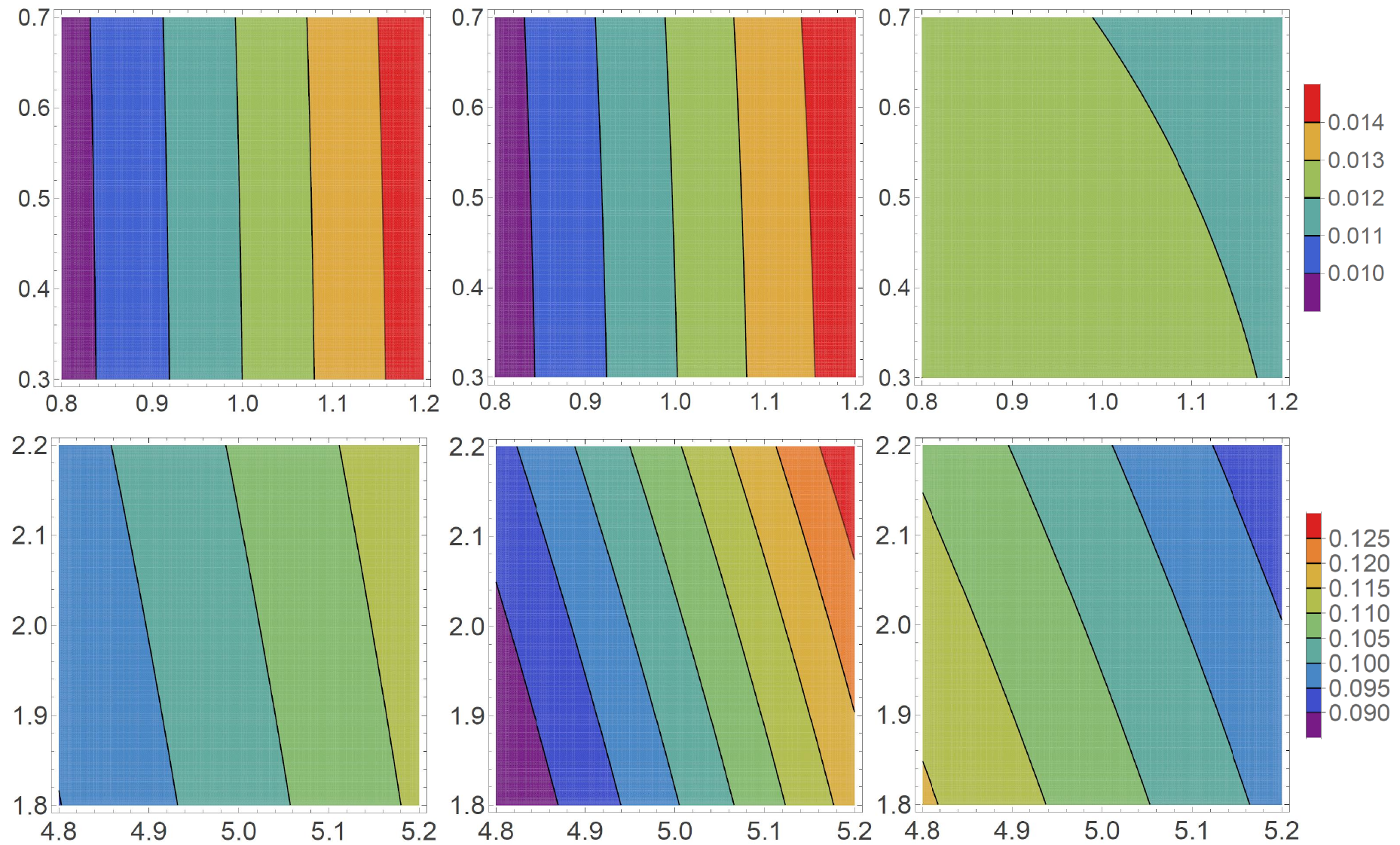}
\caption{The fields $T^{0x}(\vec r)$ (left), $T^{0x}(\vec r)/T^{00}(\vec r)$ (center), and $T^{00}(\vec r)$ (right). In the top row the center of the domain is the point $(1.0,0.5)$ and in the bottom row the center is the point $(5.0,2.0)$. In each row the data in the second two figures are scaled to match the first at the central point.} 
\label{plot-eng}
\end{figure}

The similarity between $\omega^z(\vec r)$ and $L^z_{\rm mod}(\vec r)$ motivates us to look directly at the different contributions to the integrands in these calculations. In Fig.~\ref{plot-eng} we show $T^{0x}(\vec r)$ (left), $T^{0x}(\vec r)/T^{00}(\vec r)$ (center), and $T^{00}(\vec r)$ (right) computed for $b=2$~fm. The top row shows these functions in a small domain around the point $(1.0, 0.5)$~fm and the bottom row is centered at the point $(5.0,2.0)$~fm. Within each row the second two fields have been scaled so they match the first field at the center of the domain. The top row of the figure shows that close to the center of the collision the energy density is approximately constant and the velocity behaves similarly to the Poynting vector. At the point $(5.0,2.0)$~fm, where the value of $L^z(\vec r)$ is large and positive, the gradients of the energy density are not small and the Poynting and velocity fields are not close to each other. Thus we see that one source of the difference between $L^z(\vec r)$  and $\omega^z(\vec r)$ is the spatial inhomogeneity of the energy density field.

In addition to the difference between the velocity vector and the Poynting vector, there is another important factor responsible for the different behaviour of  $L^z(\vec r)$  and $\omega^z(\vec r)$. As discussed in Sec.~\ref{sec-fourier}, there is a striking similarity between the structure of the vorticity (\ref{model-vort}) and the leading term in the expression for the local angular momentum (\ref{leading-L}) which is proportional to $\Delta^4$. However the similar structure of these two expressions is misleading. In fact the $\Delta^4$ term in Eq.~(\ref{leading-L}) is exactly zero. To prove this statement we use the equation of universal flow which was introduced in \cite{Vredevoogd:2008id} and studied in detail in context of glasma in our work \cite{Carrington:2024utf}. The equation of universal flow states that at mid-rapidity the components of the Poynting vector are related to the gradient of the energy density as
\be
T^{0x}(t,\vec r) = -\frac{1}{2} t \frac{\partial T^{00}(t,\vec r)}{\partial x},
~~~~~~~~~~
T^{0y}(t,\vec r) = -\frac{1}{2} t \frac{\partial T^{00}(t,\vec r)}{\partial y},
\label{uni-flow}
\ee
where $t$ is the time which coincides with the proper time at mid-rapidity. These equations can be derived as approximate relations for boost invariant systems, but in \cite{Carrington:2024utf} we showed that they are exactly satisfied by the proper time expanded energy-momentum tensor up to seventh order. Using Eq.~(\ref{uni-flow}) it is easy to see that the term of order $\Delta^4$ in Eq.~(\ref{modelLz}) is identically zero. The behaviour of $L^z(\vec r)$ is therefore dominated by the term of order $\Delta^6$ in Eq.~(\ref{modelLz}). To verify that $L^z(\vec r)/\Delta^6$ is approximately independent of $\Delta$, we have calculated $L_z$ for $b=2$~fm and two values $\Delta=0.20$ and $\Delta=0.15$. The relative difference of the average of the absolute value of the results across the transverse plane is $2.9\%$.

We emphasize that the vanishing of the contribution of order $\Delta^4$ to the local angular momentum means that when $\Delta$ is small the dominant contribution to $L^z(\vec r)$ comes from the third spatial derivatives of the Poynting vector. The vorticity is given by the first derivatives, which explains the qualitatively different behaviour of these two quantities. We also note that the equation of universal flow is valid to high accuracy not only for glasma but for any system for which the energy-momentum tensor is boost invariant and mostly diagonal. The vanishing of the contribution to $L^z(\vec r)$ that is proportional to the first spatial derivative of the momentum flow is therefore a general feature of ultrarelativistic systems.

\begin{figure}[t]
\includegraphics[width=17cm]{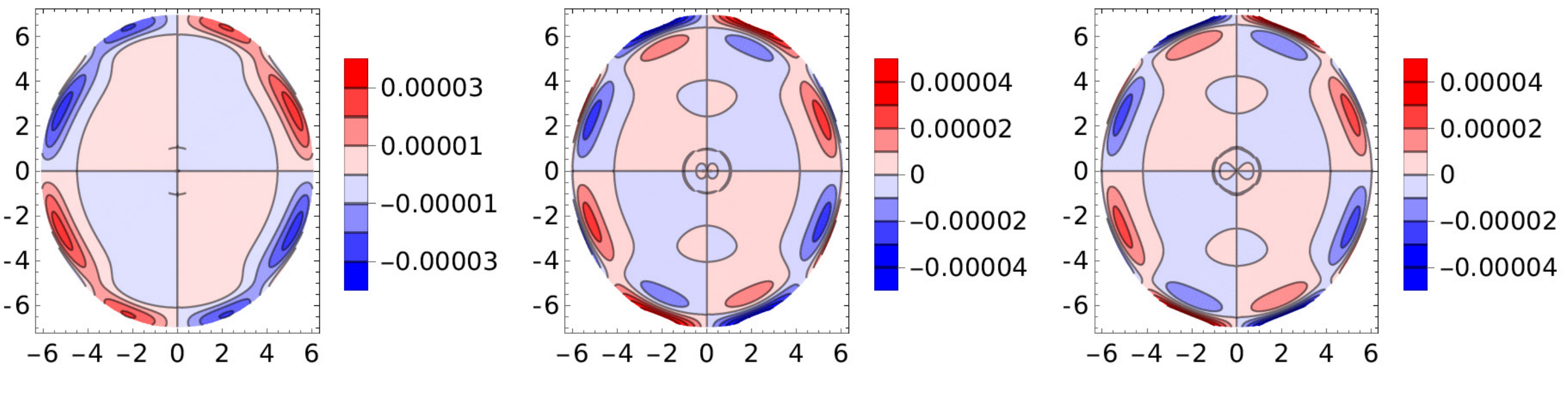}
\includegraphics[width=17cm]{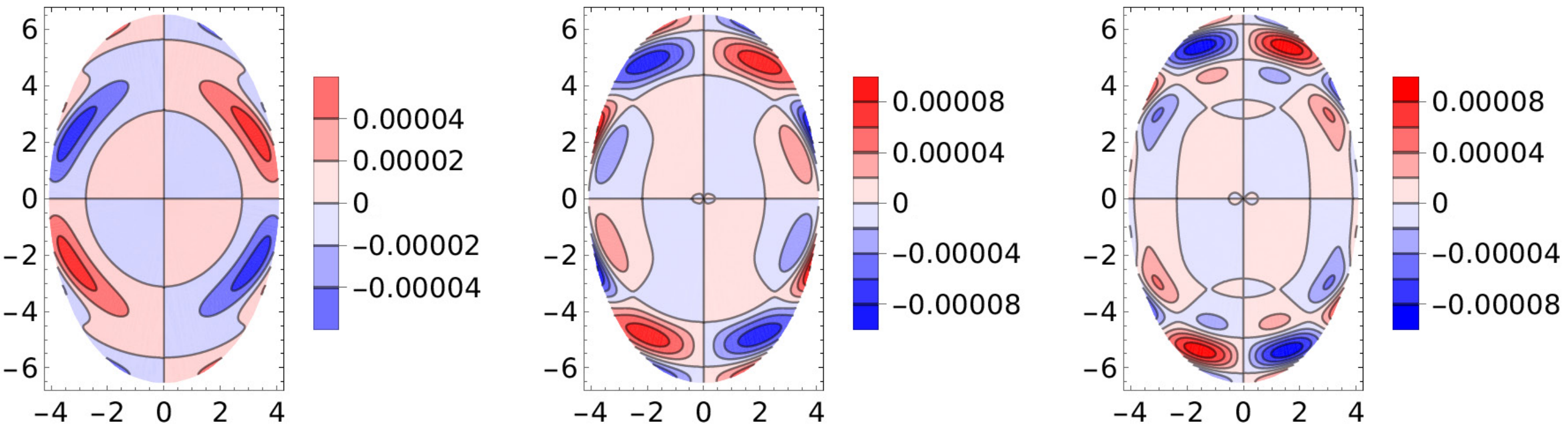}
\vspace{-7mm}
\caption{Fourier decomposition of $L^z(\vec r)$ for $b=2$~fm (top row) and $b=6$~fm (bottom row) showing only terms of order $\Delta^6$. The Fourier components $n=0$ are shown in the left column, the sum of Fourier components $n=0,2$ and $n=0,2,4$ are in the center and right columns, respectively.}
\label{D46b2-b6}
\end{figure} 

Finally, we check how well we can reproduce the glasma local angular momentum with the first three Fourier harmonics which contribute to the term proportional to $\Delta^6$ from Eq.~(\ref{modelLz}). In Fig.~\ref{D46b2-b6} we show the Fourier decomposition of this term for $b=2$~fm (top row) and $b=6$~fm (bottom row). The sum of the Fourier components $n=0$, $n=0,2$ and $n=0,2,4$ are shown in the left, center and right columns, respectively. Comparing the plots from the right column of Fig.~\ref{D46b2-b6} to those from Fig.~\ref{plotcc}, one finds that the term of order $\Delta^6$ in Eq.~(\ref{modelLz}) reproduces the glasma local angular momentum reasonably well. The upper row of Fig.~\ref{D46b2-b6} shows that at small impact parameter an octupole contribution develops when harmonics $n>0$ are added to the expansion. The lower row of the figure shows that at large impact parameter the octupole structure is barely seen even when the harmonic $n=4$ is taken into account. This agrees with the impact parameter dependence of the quadrupole and octupole contributions to $L^z(\vec r)$  shown in Fig.~\ref{plotLz-bvals}.
 
The results presented in Secs.~\ref{sec-vorticity} and \ref{sec-Lz} demonstrate that the glasma vorticity and local angular momentum along the beam axis are qualitatively different. The analysis in this section explains the origin of the differences. We comment that our findings are not surprising if one considers the elementary argument leading to the conclusion that vorticity and angular momentum should be closely related. That argument is as follows. The angular momentum of a rigid body uniformly rotating around a fixed axis with angular velocity $\vec\Omega$ is $\vec L = I \vec\Omega$ where $I$ is the moment of inertia. The vorticity of a rotating body in a reference frame at rest equals $\vec{\omega} = 2 \vec{\Omega}$. One therefore has $\vec L = I \vec\omega/2$. The glasma however is very far from a uniform rigid body, which means that this simple argument cannot be safely applied to glasma, and that the vorticity and local angular momentum of glasma can be very different from each other.

%%%%%%%%%%%%%%%%%%%%%%%%%%%%%%%%%%%%%%%%%%%%%%%%%%%%%%
\section{Relevance of our results}
\label{sec-analysis}
%%%%%%%%%%%%%%%%%%%%%%%%%%%%%%%%%%%%%%%%%%%%%%%%%%%%%%

As explained in Sec.~\ref{sec-method}, our calculation is done using a proper time expansion. Because of compute time and memory constraints we are limited to eighth order in the expansion. At this order the radius of convergence is $\tau \lesssim 0.08$~fm. For this reason all the results presented in this paper are calculated at the very early time $\tau =0.06$~fm. Since our method only works at early times, it is unclear if our results have any relevance for a description of the later stages of the collision. 

In Refs.~\cite{Carrington:2021qvi,Carrington:2023nty,Carrington:2024utf} we showed that some hydrodynamic features of the quark-gluon plasma in local equilibrium are already present in far from equilibrium glasma at very early times. In Sec.~\ref{sec-vorticity} we have shown that the glasma vorticity is another example of a quantity that is qualitatively similar to that from hydrodynamic calculations of the late stage of heavy-ion collisions. In general, this early time hydrodynamic-like behaviour occurs because the glasma energy-momentum tensor satisfies the equations of universal flow (\ref{uni-flow}) that give rise to the glasma fluid-like behaviour, as discussed in detail in Ref.~\cite{Carrington:2024utf}. These considerations justify the expectation that our results \sm{may} have implications for understanding experimental observations of the polarization of final-state hadrons.

It is interesting to compare the magnitude of the vorticity and local angular momentum we have computed. A direct comparison is not possible because vorticity has dimension inverse length and angular momentum is dimensionless.  To make a meaningful comparison we consider the local angular momentum computed in a box of  size $\Delta$ with the vorticity multiplied by $\Delta$. We find that the average of the absolute values of $L^z(\vec r)$ at $b=2$~fm is equal to the average of $|\omega^z(\vec r)| \Delta$ at the same impact parameter if $\Delta \approx 0.4$~fm. This means that the magnitude of the two quantities is roughly the same  at a scale close to a hadron radius.

The next question to address is whether either $L^z(\vec r)$ or $\omega^z(\vec r)$  is big enough to generate the observable polarization effect. To perform an order magnitude quantitative analysis, we will use the only widely accepted and commonly applied approach, which assumes that the matter from which final-state hadrons originate is in a state of global thermodynamic equilibrium. In this case the magnitude of the polarization can be roughly estimated by considering a non-relativistic system in thermodynamic equilibrium which rotates with angular velocity $\vec{\Omega}$, see \S~34 of the textbook \cite{Landau-Lifshitz-SM-1980}. The Gibbs distribution of the angular momentum $\vec{j}$ of a single particle, which splits into spin $\vec{s}$ and orbital angular momentum $\vec{l}$ as $\vec{j}= \vec{l} + \vec{s}$, is proportional to $\exp\big(- \vec{\Omega} \cdot \vec{j}/T\big)$ where $T$ is the system's temperature. The vorticity of the rotating system computed in a reference frame at rest and defined by Eq.~(\ref{vort-def}) equals $\vec{\omega} = 2 \vec{\Omega}$, so the distribution of spin projections on the quantization axis along the vector $\vec{\omega}$ is proportional to $\exp\big(-\omega s/(2T)\big)$. Since hyperons are spin 1/2 particles one  finds the polarization $P$ as
\be
P \equiv  \frac{N_+ - N_-}{N_+ + N_-} = \frac{e^{\frac{\omega}{4T}} - e^{-\frac{\omega}{4T}}}{e^{\frac{\omega}{4T}} + e^{-\frac{\omega}{4T}}}
~~ \underset{\omega \ll T}{\approx} ~~ \frac{\omega}{4T} ,
\ee
where $N_{\pm}$ are the numbers of particles with  spin parallel and antiparallel to $\vec{\omega}$. For a relativistic treatment of the problem see the reviews \cite{Florkowski:2018fap,Becattini:2024uha}. In the measurements of Refs.~\cite{STAR:2019erd,STAR:2023eck,ALICE:2021pzu} a polarization of the order of one per mille was obtained. To generate such a polarization the quantity $\omega/(4T)$ should be of order  $10^{-3}$ at the freeze-out stage of relativistic heavy-ion collisions. Since the freeze-out temperature is about 160 MeV, the vorticity must be of order $\omega \sim 10^{-3}~{\rm fm}^{-1}$. To get a crude estimate of the corresponding value of the  glasma vorticity, we treat the matter produced in a relativistic heavy-ion collision as an ideal fluid. In this case the integral $\int_S d^2\vec{\sigma} \cdot \vec{\omega}$  over a surface $S$ is conserved, see \S~8 of the textbook \cite{Landau-Lifshitz-FM-1987}. This means that the vorticity moves with the fluid and it is conserved if the surface $S$ is conserved. If we assume that the vorticity of the strongly interacting matter is approximately conserved, then in order to conclude that the observed $\Lambda$ polarization could have its origin in the glasma we would need a glasma vorticity of approximately $10^{-3}~{\rm fm}^{-1}$. The magnitude of the glasma vorticity shown in Fig.~\ref{fig-vort} is of order $10^{-3}~{\rm fm}^{-1}$ only in the outer part of the interaction region where the distance from the center of the interaction region is approximately 5.0~fm to 6.0~fm.\footnote{We remind the reader that our calculations become unreliable at large transverse distances because of the gradient expansion we use.} Although this region appears to be small, its surface area is about 40\% of the transverse area of the interaction region. It is therefore reasonable to expect that the vorticity of the glasma  could lead to an observable polarization effect. The question, however, certainly needs to be studied further.

%%%%%%%%%%%%%%%%%%%%%%%%%%%%%%%%%%%%%%%%%%%%%%%%%%%%
\section{Discussion, summary and conclusions}
%%%%%%%%%%%%%%%%%%%%%%%%%%%%%%%%%%%%%%%%%%%%%%%%%%%%

We have calculated vorticity and local angular momentum of the glasma from the earliest stage of ultrarelativistic heavy-ion collisions. Contrary to expectation, $L^z(\vec r)$ and $\omega^z(\vec r)$ are qualitatively different. It is therefore important to understand which of these quantities is relevant to the polarization of final-state hadrons. 

Hydrodynamic approaches based on the assumption of local thermodynamic equilibrium, reviewed in \cite{Florkowski:2018fap,Becattini:2024uha}, use the thermal vorticity defined by Eq.~(\ref{thermal-vort}) to compute the hadron polarization. The sign of the obtained vorticity, which agrees with ours, is opposite to the sign of the measured polarization, see e.g. \cite{Becattini:2021iol}. Hydrodynamic calculations can produce the correct sign, when what are called shear effects are included, but the method to include these effects is not well established. One should remember that a hydrodynamic description requires that thermodynamic parameters like the fluid four-velocity and temperature depend only weakly on the spatial variables, so that the zeroth and low order derivatives are sufficient to characterize the system. The requirement of weak inhomogeneity is consistent with the original idea that vorticity controls a particle's polarization, which comes from an analysis of the rigid rotation of a system in global thermal equilibrium (see Sec.~\ref{sec-compare}). However, it is unclear whether the requirement of weak inhomogeneity is satisfied by the system produced in relativistic heavy-ion collisions. 

It should be also stressed that the assumption that the spin degrees of freedom are in thermodynamic equilibrium, which is usually adopted in a hydrodynamic formulation, is difficult to justify. Collisions of particles that involve a spin flip are usually less probable than those that conserve spin, and spin equipartition is typically much slower than energy equipartition \cite{Kapusta:2019sad,Kapusta:2020npk,Ayala:2020ndx,Hidaka:2023oze,Wagner:2024fhf}. This means that the spin degrees of freedom in heavy-ion collisions probably do not reach the equilibrium by the time of freeze-out. The polarization of final-state hadrons is then a consequence of the dynamical evolution of spin degrees of freedom constrained by the conservation of angular momentum. A method that uses this approach was initiated in \cite{Florkowski:2017ruc}, see also \cite{Florkowski:2018fap}, and is called ideal spin hydrodynamics. Recently dissipative effects have been taken into account \cite{Sapna:2025yss}. 

This discussion raises another potentially confusing issue. One could object to our conclusions by arguing that if the equilibration of spin degrees of freedom is slow because the spin-flip processes are rare, then the system's total angular momentum would be dominated by the orbital angular momentum at any time due to its weak coupling to particle spins. However, it is important to remember that the vast majority of the particles observed in relativistic-heavy-ion collisions are produced through hadronization, and since total angular momentum is conserved, it is natural to expect that spin of produced particles is correlated with the angular momentum of the incoming particles. Since the observed polarization is of order ${\cal O}(10^{-3})$ such a mechanism is not excluded. 

In line with the above discussion our results also suggest that local angular momentum is a more relevant characteristic than either vorticity or thermal vorticity for determing the polarization of final-state hadrons. We first note that the sign of the quadrupole contribution to the local angular momentum, which dominates at larger impact parameters, agrees with the sign of the experimentally observed $\Lambda$ polarization. By performing a direct calculation of what we called modified $L^z(\vec r)$ (see Sec.~\ref{sec-compare}) we have verified that the difference between angular momentum and vorticity comes from sizable gradients of the energy density. We also note that the vanishing of the contribution of order $\Delta^4$ to the local angular momentum (see Sec.~\ref{sec-compare}) means that the leading contribution to $L^z(\vec r)$ comes from the third spatial derivatives of the Poynting vector. The vorticity in contrast is given by the first derivatives and therefore cannot reproduce the behaviour of $L^z(\vec r)$. Finally, we comment that although we have argued that the local angular momentum might be a useful quantity to study in order to understand the polarization of final-state hadrons, it is important to remember that a quantitative approach which would allow one to relate the local angular momentum to polarization needs to be developed.

We summarize our results and conclusions. We have studied the angular momentum of the glasma at nonzero impact parameter using a CGC approach and a proper time expansion. We have calculated the global glasma angular momentum perpendicular to the reaction plane. It is much smaller than the angular momentum of the participants of the colliding nuclei. This means that the initial angular momentum which is mostly carried by valence quarks is not transferred to the glasma. The conclusion is that the idea of a spinning fireball is not relevant for ultrarelativistic collisions. This is consistent with the fact that the observed global polarization at LHC energies is essentially zero. 

Our main focus has been the glasma vorticity and local angular momentum along the beam axis. At small impact parameters ($b \lesssim 3$~fm) the vorticity of the glasma has a quadrupole structure, but its sign is opposite to that of the simple argument based on the elliptic flow gradient which is illustrated in the left panel of Fig.~\ref{fig-argument}, and also opposite to the sign of the measured $\Lambda$ polarization along the beam axis \cite{STAR:2019erd,STAR:2023eck,ALICE:2021pzu}. It does agree with the pattern and sign of the vorticity obtained within hydrodynamical models of relativistic heavy-ion collisions \cite{Becattini:2021iol}. This is consistent with our earlier finding \cite{Carrington:2024utf} that some features of the glasma strongly resemble the fluid-like system from the later stages of the collision even though glasma dynamics is governed by Yang-Mills and not hydrodynamic equations. We have demonstrated that the quadruple structure of the vorticity is not caused by elliptic flow, as commonly believed, but is due to  non-radial flow which is bigger in the reaction plane than out of the plane. At larger impact parameters ($b \gtrsim 4$~fm), there is a superposition of quadrupole and octupole contributions to the vorticity.

The local angular momentum is qualitatively different from the vorticity. It is predominately octupole at small impact parameters ($b \lesssim 4$~fm) and quadrupole for large impact parameters ($b \gtrsim 4$~fm). This is consistent with the simple argument illustrated in the left panel of Fig.~\ref{fig-argument} which works better for large impact parameters when the elliptic flow is stronger. The sign of the quadrupole contribution to the local angular momentum agrees with the sign of the experimentally observed $\Lambda$ polarization. This observation together with the more theoretical arguments discussed earlier in this section suggest that it is the local angular momentum and neither the vorticity nor the thermal vorticity that determines the polarization of final-state hadrons. Our work motivates further effort towards the development of a quantitative approach that allows one to connect the local angular momentum to polarization.

%-----------------------------------------------------------------------
\section*{Acknowledgments}
%-----------------------------------------------------------------------

This work was partially supported by the Natural Sciences and Engineering Research Council of Canada under grant SAPIN-2023-00023. 

%\newpage

\end{document}